\documentclass[amssymb,aps,pre,letterpaper,twocolumn,showpacs]{revtex4-1}
\usepackage{graphicx}

\newif\ifpsfrag
\psfragtrue
\ifpsfrag
  \usepackage{psfrag}
\fi

\begin{document}
\def\be{\begin{equation}}
\def\ee{\end{equation}}
\def\ba{\begin{eqnarray}}
\def\ea{\end{eqnarray}}

\title{Griffiths phase and critical behavior of the
2D Potts models with long-range correlated disorder}

\author{Christophe Chatelain}
\affiliation{
Groupe de Physique Statistique,
D\'epartement P2M,
Institut Jean Lamour (CNRS UMR 7198),
Universit\'e de Lorraine, France}

\date{\today}

\begin{abstract}
The $q$-state Potts model with a long-range correlated disorder is studied
by means of large-scale Monte Carlo simulations for $q=2,4,8$ and 16. Evidence
is given of the existence of a Griffiths phase, where the thermodynamic
quantities display an algebraic Finite-Size Scaling, in a finite range of
temperatures. The critical exponents are shown
to depend on both the temperature and the exponent of the algebraic decay
of disorder correlations, but not on the number of states of the Potts model.
The mechanism leading to the violation of hyperscaling relations is observed
in the entire Griffiths phase.
\end{abstract}

\maketitle

\section{\label{Sec1}Introduction}
It is well known that the presence of impurities can greatly affect the
properties of a physical system, especially when the latter undergoes a
phase transition. In the following, the case of frozen impurities, i.e.
quenched disorder, coupled to the energy density of the system is considered.
It is assumed that no frustration is induced by randomness.
With these assumptions, Harris analyzed the conditions for a change of
critical behavior at a second-order phase transition upon the introduction
of disorder~\cite{Harris74}. In the language of Renormalization Group (RG),
disorder is a relevant perturbation when its fluctuations in a finite
domain grow faster than the fluctuations of the energy, or equivalently,
when the specific heat exponent $\alpha$ of the pure model is positive.
% Experimental examples are numerous (??!)
On the theoretical side, the $q$-state Potts model provides a useful and
simple toy model to study the influence of disorder. In two dimensions,
the pure model undergoes a second-order phase transition when $q\le 4$
with a $q$-dependent universality class. Upon the introduction of quenched
disorder, the critical behavior is governed by a new $q$-dependent RG
fixed point when $2<q\le 4$. A good agreement between numerical
calculations~\cite{RBPM-MC} and RG series expansions~\cite{RBPM-RG}
of the critical exponents has been achieved.
In the case $q=2$, equivalent to the Ising model, disorder is a marginally
irrelevant perturbation and the critical behavior is only modified by
multiplicative logarithmic corrections~\cite{RBIM}. In the three-dimensional
case, only the Ising model undergoes a second-order phase transition.
As predicted by the Harris criterion, a new critical behavior is
found by RG studies and observed numerically~\cite{RBIM3D}.
\\

When the pure system undergoes a first-order phase transition, the
introduction of disorder softens the energy jump and reduces the
latent heat~\cite{ImryWortis}. For a strong enough disorder, a continuous
phase transition can be induced. In two dimensions, an infinitesimal amount
of disorder is sufficient to induce a second order phase
transition~\cite{HuiBerker,Aizenman}. This rigorously-proved
statement was first tested in the case of the 8-state Potts
model~\cite{Landau}. In the entire regime $q>4$, where a first-order phase
transition is undergone by the pure Potts model, a continuous transition
is induced with a $q$-dependent critical behavior~\cite{RBPotts}.
In three dimensions, the softening of the transition was observed
numerically for the $q=3$~\cite{RBPM3-3D} and $q=4$~\cite{RBPM-3D}
Potts models, as well as in the limit $q\rightarrow +\infty$~\cite{RBPMInf-3D}.
For weak disorder, the transition remains discontinuous, while at strong
disorder, it becomes continuous.
\\

It was implicitly assumed in the above discussion that the disorder
was uncorrelated. This may not be the case anymore if the impurities
interact and were given enough time to equilibrate. This situation
may be encountered with charged impurities. At a second-order
phase transition, long-range correlated disorder leads to a critical
behavior that can be distinct from the case of short-range or
uncorrelated disorder. Weinrib and Halperin studied by RG the
$n$-component $\phi^4$-model in dimension $d$ with a correlated disorder
decaying algebraically with an exponent $a$~\cite{Weinrib}.
In the same spirit as the Harris criterion, such a long-range correlated
disorder is shown to be relevant when the correlation exponent $\nu$ of
the pure model satisfies the inequality $\nu<2/a$. Interestingly, disorder
is a marginally irrelevant perturbation at the new long-range random
fixed point, which means that $\nu=2/a$. This relation was proved to be
exact at all orders in perturbation~\cite{Honkonen} and was later
confirmed by Monte Carlo simulations of the 3D Ising
model~\cite{IMCorrele} and an explicit RG calculation of
the 2D Ising model~\cite{Rajabpour}.
\\

Weinrib and Halperin calculation is based on the assumption that 
disorder correlations are isotropic and that $n$-point disorder
cumulants are irrelevant for $n>2$. Anisotropically correlated
disorder is therefore out of its range of validity. The latter
has attracted a lot of attention since the introduction of the
celebrated McCoy-Wu model~\cite{McCoyWu}, which corresponds to an
Ising model with randomly distributed couplings $J_1$ and $J_2$
in one direction and infinitely correlated in the second direction
of the square lattice~\cite{McCoyWu}.
While planar defects lead to a smearing of the transition of
the 3D Ising model~\cite{Vojta04}, the phase transition of this
2D Ising model with parallel linear defects remains sharp.
Exploiting the mapping to the Ising quantum chain in a transverse
field, the critical exponents %of the McCoy-Wu model
were determined exactly: $\beta=(3-\sqrt 5)/2$ and $\nu=2$~\cite{DFisher95}.
Because of the layered structure, the anisotropy exponent $z$
is infinite. The isotropy can be restored by an appropriate superposition
of two McCoy-Wu models oriented in two different directions. A different
magnetic critical behavior is then observed~\cite{Turban}.
When algebraically decaying disorder correlations are introduced
in the transverse direction of the McCoy-Wu model, the Weinrib-Halperin
law $\nu=2/a$, where $a$ is the disorder correlation exponent,
is recovered~\cite{Rieger99}.
Interestingly, the critical behavior of the Potts model with
uncorrelated homogeneous disorder was conjectured to be described
by an isotropic version of the McCoy-Wu-Fisher fixed point
in the limit $q\rightarrow +\infty$~\cite{Mercaldo}.
\\

The McCoy-Wu model is easily extended to $q$-state Potts
spins. In the regime $q\le 4$, the critical behavior was proved to be
independent of the number of states $q$ and therefore identical to
that of the original McCoy-Wu model~\cite{Senthil}. Numerical
calculations showed  that the first-order phase transition of the Potts
model with $q>4$ is completely rounded, like in the case of homogeneous
disorder, but, in contrast to homogeneous disorder, the critical behavior
induced by disorder is independent of $q$ and therefore described by the
infinite-disorder McCoy-Wu-Fisher fixed point~\cite{Carlon}. Different
arrangements of correlated couplings were also studied. An aperiodic
sequence of couplings $J_1$ and $J_2$ in one direction, infinitely
correlated in the second direction of a square lattice, also provokes
the rounding of the first-order phase transition of the Potts model
when the wandering exponent of the sequence is sufficiently
large~\cite{Aperiodic}. However, in contrast to the McCoy-Wu model, the
induced critical behavior depends on the number of states $q$.
\\

The McCoy-Wu model is the first model where a Griffiths phase was observed~%
\cite{Griffiths}. In a finite range of temperatures around the critical
point, the free energy is a singular function of the magnetic field.
Consequently, the magnetization is also singular and the magnetic
susceptibility diverges for all temperatures in the Griffiths phase.
The Griffiths phase should also be present in the aforementioned Potts
models with isotropic disorder but it is believed to be too weak to be
observed numerically. In the following, the mechanism leading to such a
Griffiths phase is briefly discussed in the case of a binary distribution
of strong and weak couplings, say $J_1$ and $J_2$ respectively.
% The case of dilution is recovered by taking $J_1=J$ and $J_2=0$???. 
The critical temperature $T_c$ of the random system is expected to lie
between the critical temperatures $T_1$ and $T_2$ of homogeneous systems
with only strong or weak couplings. A rough estimate of these temperatures
is given by $k_BT_1\simeq J_1$ and $k_BT_2\simeq J_2$. In the paramagnetic
phase $T>T_c$, the Griffiths phase is caused by the existence of
rare macroscopic regions with a high concentration of strong couplings $J_1$.
While the rest of the system is still not ordered, they can
order independently for all temperatures $T_c\le T<T_1$~\cite{Vojta}.
Because these regions are macroscopic, they cannot be easily flipped by
local Monte Carlo algorithms. The spin-spin autocorrelation functions
can be shown to decay much slower than the expected exponential decay
in a paramagnetic phase. Below the critical temperature, the Griffiths
phase is caused by the existence of ordered domains isolated of the rest
of the system by a high concentration of weak bonds at its frontiers.
This mechanism is therefore effective for temperatures $T_2\le T\le T_c$.
\\

Recently, we have studied the 2D 8-state Potts model with an isotropic
correlated non-Gaussian disorder~\cite{EPL}. Such a disorder was obtained by
simulating an Ashkin-Teller model on a critical line of its phase diagram.
To each independent spin configuration is assigned a coupling configuration
for the Potts model. We have shown by Monte Carlo simulations that the
first-order phase transition of the 8-state Potts model is rounded.
Disorder fluctuations are large and cause the violation of the hyperscaling
relation, in the same way as in the 3D random-field Ising model, even though
no frustration is present. In this work, we show by means of
large-scale Monte Carlo simulations that this behavior is actually observed
in a finite range of temperatures and for all numbers of states $q$ of
the Potts model. The equivalent of 25 years on a single CPU was used.
The paper is organized as follows. Details about the model and the
simulations are given in section~\ref{Sec2}.
The phase diagram is discussed and evidences are given of the existence
of a Griffiths phase in section~\ref{Sec3}. The critical behavior
in this Griffiths phase is then presented in section~\ref{Sec4}.
Non self-averaging properties and hyperscaling
violation are finally discussed in section~\ref{Sec5}.
A conclusion follows.

\section{\label{Sec2}Models and numerical details}
\subsection{The Potts model}
The Potts model is a generalization of the Ising model where the
classical degrees of freedom, called spins, lie on the nodes
of a lattice and can take $q$ possible values, for example
$\{0,1,\ldots,q-1\}$. In the following, $\sigma_i$ denotes the spin
lying on the $i$-th node of the lattice. Spins are assumed to interact
only if they are on nearest-neighboring sites. The Potts Hamiltonian
is defined as~\cite{Potts}
   \be H=-\sum_{(i,j)} J\delta_{\sigma_i,\sigma_j}
   \label{HPotts}\ee
where the sum is restricted to pairs of nearest neighbors $i$ and $j$
of the lattice. The Ising model is recovered in the case $q=2$.
Note that the Fortuin-Kasteleyn representation of the Potts model allows
for a generalization to non-integer values of $q$~\cite{Fortuin}.
As mentioned in the introduction, the Potts model undergoes a
phase transition which is of first order for $q>q_c$ and continuous
for $q\le q_c$. The value $q_c$ depends on the number of dimensions
of the lattice. In the two-dimensional case, $q_c=4$ and duality
arguments based on the Fortuin-Kasteleyn representation show that
the transition temperature is given by
     \be \big(e^{\beta J}-1\big)=\sqrt q\ee
where $\beta=1/k_BT$.
The Potts Hamiltonian (\ref{HPotts}) is unchanged under the global
circular shift $\sigma_i\rightarrow \sigma_i=(\sigma_i+1)\ {\rm mod}\ q$.
This ${\mathbb Z}_q$-symmetry is broken in the low-temperature phase,
where a majority of spins can be found in the same state. The order
parameter $m$, called magnetization, can therefore be defined as
   \be m={q\rho_{\rm max.}-1\over q-1}\ee
where $\rho_{\rm max.}$ is the fraction of spins in the majority state.
\\

In the following, we are interested in the influence of quenched disorder
coupled to the energy density. On each lattice edge $(i,j)$, joining the two
neighboring sites $i$ and $j$, the exchange coupling $J$ appearing in
the Hamiltonian (\ref{HPotts}) is replaced by a random variable $J_{ij}$:
   \be H=-\sum_{(i,j)} J_{ij}\delta_{\sigma_i,\sigma_j}
   \label{H_RBPM}\ee
The calculation of the thermodynamic quantities requires an average
over both the thermal fluctuations and the coupling configurations:
   \be \overline{\langle X\rangle}=\sum_{\{J_{ij}\}} \wp(\{J_{ij}\})
   \left[{1\over{\cal Z}[J_{ij}]}\sum_{\{\sigma_i\}} e^{-\beta H[\sigma_i,
       J_{ij}]}\right].\ee
The average over thermal fluctuations is denoted by brackets, for example
$\langle m\rangle$ in the case of magnetization. The over line corresponds
to the average over the probability distribution $\wp(\{J_{ij}\})$ of the
random couplings $J_{ij}$. Duality arguments can be extended to the
random case. Consider a coupling configuration $\{J_{ij}\}$. The image of
this configuration under the duality transformation is a new coupling
configuration $\{J_{ij}^*\}$ where the dual bonds are defined by~\cite{Kinzel}
   \be \big(e^{\beta J}-1\big)\big(e^{\beta J^*}-1\big)=q.\ee
The singular part of the average free energy density is unchanged if the
self-duality condition
   \be\wp(\{J_{ij}\})=\wp(\{J_{ij}^*\})  \label{SelfDuality1}\ee
on the probability distribution of the random couplings is satisfied.
In numerical studies, the binary distribution
    \be \wp(J_{ij})={1\over 2}\delta(J_{ij}-J_1)
    +{1\over 2}\delta(J_{ij}-J_2) \label{Binary}\ee
is usually the simplest choice. The critical line is therefore
given by the self-duality condition
   \be J_1=J_2^*\ \Leftrightarrow\ 
   \big(e^{\beta_c J_1}-1\big)\big(e^{\beta_c J_2}-1\big)=q.
   \label{SelfDual2}\ee
Note that the self-duality condition (\ref{SelfDuality1}) is a constraint
on the probability distribution $\wp(\{J_{ij}\})$. In the case of the
binary distribution (\ref{Binary}), it does not impose the strict equality
of the numbers of couplings $J_1$ and $J_2$ in each disorder realization,
though these numbers will converge towards the same value in the
thermodynamic limit. In the framework of RG, the fluctuations of the
numbers of couplings $J_1$ and $J_2$ in a coupling configuration
have been moreover shown to be an irrelevant perturbation at the
random fixed point~\cite{AharonyHarris}.
The self-duality condition (\ref{SelfDuality1}) is very general and
still holds in the case of correlated disorder.

\subsection{The auxiliary Ashkin-Teller model}
In order to generate the coupling configurations $\{J_{ij}\}$ of the
Potts model, an auxiliary isotropic Ashkin-Teller model is
simulated. This model corresponds to two Ising models locally coupled by their
energy density and is defined by the Hamiltonian~\cite{AshkinTeller}
   \be -\beta H^{\rm AT}=\sum_{(i,j)} \big[J^{\rm AT}\sigma_i\sigma_j
   +J^{\rm AT}\tau_i\tau_j+K^{\rm AT}\sigma_i\sigma_j\tau_i\tau_j\big].\ee
This Hamiltonian is invariant under the reversal of all spins
$\sigma_i$, all spins $\tau_i$ or both $\sigma_i$ and $\tau_i$. Two
order parameters can be defined: magnetization $M=\langle\left|\sum_i\sigma_i
\right|\rangle$ and polarization $P=\langle\left|\sum_i \sigma_i\tau_i\right|
\rangle$. The phase diagram of the Ashkin-Teller model presents several lines
separating a paramagnetic phase ($M=P=0$), a Baxter phase where all spins
are in the same state ($M,P\ne 0$) and a phase where each Ising replica is
disordered but there exists order between them ($M=0,P\ne 0$). The line
separating the paramagnetic and Baxter phases is given by self-duality
arguments~\cite{Fan}:
    \be e^{-2K}=\sinh 2J.\ee
The critical exponents along this line were obtained through the conjecture
of a mapping~\cite{Kadanoff} of the Ashkin-Teller model onto an
eight-vertex model exactly solved by Baxter. In terms of the parameter
$y\in[0;4/3]$ of the eight-vertex model and related to the couplings along
the line by
    \be\cos{\pi y\over 2}={1\over 2}\left[e^{4K}-1\right],\ee
these critical exponents read
    \be\beta_\sigma^{\rm AT}={2-y\over 24-16y},\quad
    \beta_{\sigma\tau}^{\rm AT}={1\over 12-8y},\quad
    \nu^{\rm AT}={2-y\over 3-2y}.\ee
Note that $\beta_\sigma^{\rm AT}/\nu^{\rm AT}=1/8$ is constant while
$\beta_{\sigma\tau}^{\rm AT}/\nu^{\rm AT}$ varies along the self-dual line.
Therefore, polarization-polarization correlation functions decay
algebraically
  \be\langle\sigma_i\tau_i\sigma_j\tau_j\rangle
  \sim |\vec r_i-\vec r_j|^{-a}\ee
with an exponent
  \be a=2\beta_{\sigma\tau}^{\rm AT}/\nu^{\rm AT}
  ={1\over 4-2y}\ee
that can be tuned by moving along the critical line. In order to
construct correlated coupling configurations for the Potts model,
a set of typical spin configurations of the Ashkin-Teller model at
different points of the self-dual line are first generated by
Monte Carlo simulation.
We used the cluster algorithm proposed by Wiseman and Domany~\cite{Salas}.

To each spin configuration is then associated a coupling configuration
of the Potts model by
    \be J_{ij}={J_1+J_2\over 2}+{J_1-J_2\over 2}\sigma_i\tau_i
    \label{PolAT-Coupling}\ee
where the site $j$ is located after the site $i$ on the lattice, i.e.
on its right or below. At each site, two couplings are therefore
always identical, one being horizontal and the other vertical.
Because of this construction, disorder correlations
$\overline{(J_{ij}-\bar J)(J_{kl}-\bar J)}$ decay algebraically,
like the polarization, with an exponent $a$ that can be tuned.
We have considered the six values
$y\in\{0,0.25,0.50,0.75,1,1.25\}$ corresponding to the exponents $a\simeq
0.25$, $0.286$, $0.333$, $0.4$, $0.5$ and $0.667$. The two couplings $J_1$
and $J_2$ are chosen to be related by
   \be \big(e^{J_1}-1\big)\big(e^{J_2}-1\big)=q
   \label{SeldDuality}\ee
corresponding to the self-duality condition (\ref{SelfDual2}) when
$\beta=1/k_BT=1$, i.e. the self-dual point is located at $\beta_c=1$.
At this temperature, $J_1$ is indeed equal to $J_2^*$.
Because of the construction (\ref{PolAT-Coupling}), the duality
transformation $J_{ij}\rightarrow J^*_{ij}$ is equivalent to a global
reversal $\sigma_i\tau_i\rightarrow -\sigma_i\tau_i$ of the local
polarization of the auxiliary Ashkin-Teller model. Since the Hamiltonian
of the latter is unchanged under this transformation, the Boltzmann
weight is unaffected. In terms of coupling configurations, this leads
to the self-duality condition (\ref{SelfDuality1}).
\\

On average, the total number of strong and weak couplings are equal.
However, in a finite system, these numbers fluctuate from
sample to sample. According to the central limit theorem, these
fluctuations vanish as $1/\sqrt N\sim 1/L$, where $N=L^2$ is
the total number of sites, in the case of uncorrelated disorder.
For the correlated disorder introduced above, these fluctuations
scale with the lattice size as the polarization of the Ashkin-Teller model,
i.e. as $L^{-\beta_{\sigma\tau}^{\rm AT}/\nu^{\rm AT}}=L^{-a/2}$. Since the values
$a\simeq 0.25$, $0.286$, $0.333$, $0.4$, $0.5$ and $0.667$
are considered, fluctuations decay slower than in the case of
uncorrelated disorder. Moreover, for the lattice sizes considered
in this work, these fluctuations are much smaller for uncorrelated
disorder than for correlated one. For $L=256$ and $y=0.75$, an average
polarization $\overline{|p|}\simeq 0.343(1)$ is measured while
for uncorrelated disorder, the equivalent quantity reconstructed
from the couplings is $\overline{|p|}\simeq 0.0499(1)$ for $L=16$ already.
\\

Note that, in the following, this disorder will be referred to as correlated
disorder to distinguish it from uncorrelated disorder. However, it is
important to keep in mind that, as described above, it was obtained using a
very particular construction and does not display the same properties as
the correlated disorder considered by Weinrib and Halperin. In the latter,
disorder was indeed distributed according to a Gaussian probability
distribution so that $2n$-point correlation functions are related to
two-point correlations by the Wick theorem. This is not the case for the
Ashkin-Teller model and, therefore, for the couplings that are generated
from the typical spin configurations of this model.

\section{\label{Sec3}Temperature dependence}
In this section, the temperature dependence of the average thermodynamic
quantities is investigated to determine the phase diagram of the model.
Note that the temperature affects only the Potts model and not the auxiliary
Ashkin-Teller model used to construct the coupling configurations. Disorder
correlations decay algebraically independently of the temperature. Two
numbers of Potts states, $q=2$ (equivalent to the Ising model) and $q=8$, are
considered. As mentioned in the introduction, the former undergoes a
second-order phase transition in the absence of disorder while the latter
displays a discontinuous transition. Monte Carlo simulations with the
Swendsen-Wang cluster algorithm~\cite{SW} were performed
for the lattice sizes $L=32,48,64$ and $96$. The exponent of the algebraic
decay of the disorder correlations was fixed to $a=0.4$, which corresponds
to a parameter $y=0.75$ for the auxiliary Ashkin-Teller model. For comparison,
the case of an uncorrelated disorder is also considered and presented in an
inset in each figure. The thermodynamic quantities were averaged over
$14563$ disorder configurations for $L=96$, $32768$ for $L=64$, $58254$
for $L=48$ and $131072$ for $L=32$. For each disorder configuration, 2000
Monte Carlo iterations were performed to estimate the thermal averages.
The autocorrelation time depends on the temperature but never exceeds the
values $\tau\simeq 2$ for the Ising model and $\tau\simeq 8$ for the
8-state Potts model. The case of the Ising model is first discussed.

\subsection{Ising model}
\begin{figure}
\centering
\ifpsfrag
\psfrag{<m>}[Bc][Bc][1][1]{$\overline{\langle m\rangle}$}
\psfrag{beta}[tc][tc][1][0]{$\beta$}
\psfrag{L=32}[Bc][Bc][1][0]{\tiny $L=32$}
\psfrag{L=64}[Bc][Bc][1][0]{\tiny $L=64$}
\psfrag{L=48}[Bc][Bc][1][0]{\tiny $L=48$}
\psfrag{L=96}[Bc][Bc][1][0]{\tiny $L=96$}
\setbox1=\hbox{\includegraphics[height=8.5cm]{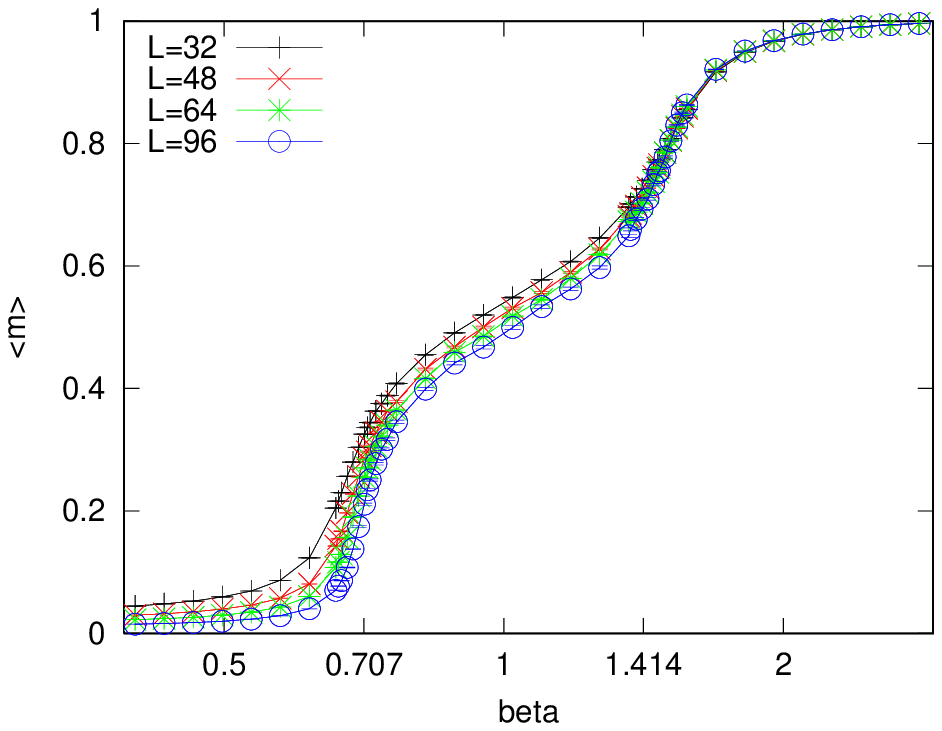}}
\includegraphics[width=8.5cm]{m-q_2-r_3-y_0.75.eps}
\llap{\makebox[\wd1][r]{\raisebox{1cm}
{\includegraphics[height=2.75cm]{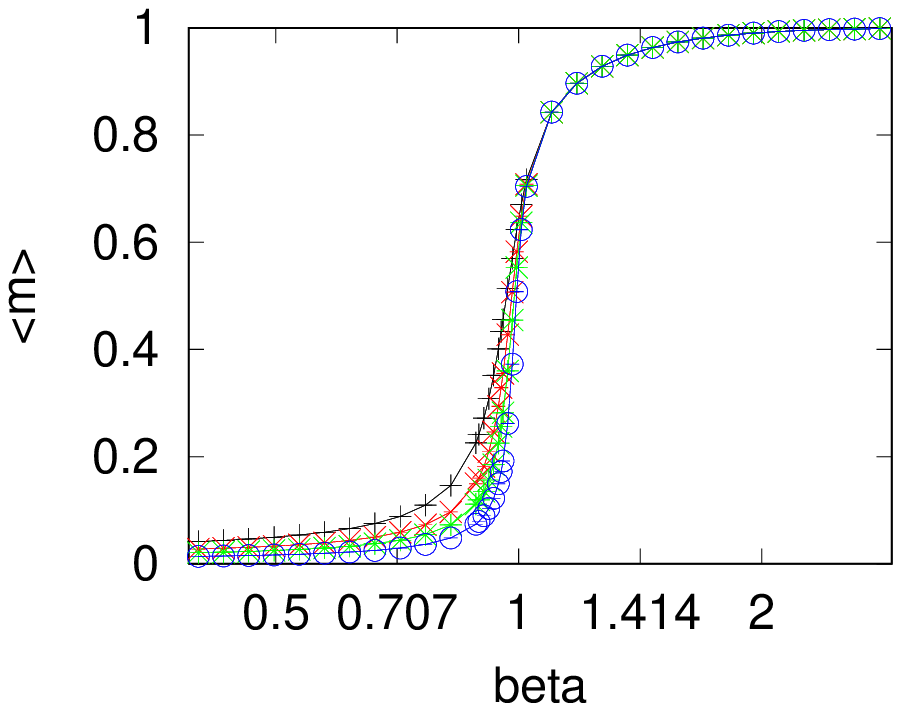}}}}
\else
\includegraphics[width=8.9cm]{Fig1.ps}
\fi
\caption{(Color online) Average magnetization of the Ising model ($q=2$), with
a disorder strength $r=3$ and a correlation exponent $a=0.4$ ($y=0.75$), versus
the inverse temperature $\beta=1/k_BT$. The different curves correspond to
different lattice sizes ($L=32,48,64$ and 96). In the inset, the magnetization
curve in the case of uncorrelated disorder is plotted.
}\label{m-q_2}
\end{figure}

As can be observed on Fig.~\ref{m-q_2}, the magnetization curve of the Ising
model with correlated disorder is not typical. Instead of the usual single
abrupt variation of magnetization, two such variations are seen. Between the
paramagnetic and ferromagnetic phases, an intermediate region of slow variation
is present. This behavior is remarkably different from the case of uncorrelated
disorder shown in the inset. With correlated disorder, the magnetization
displays a strong lattice-size dependence in the intermediate regime, similar
to what is observed in the paramagnetic phase. As will be more extensively
discussed in the next section, this behavior is algebraic. It is tempting
to associate the boundaries
of this intermediate regime to the two temperature scales introduced by the
two Potts couplings $J_1$ and $J_2$. In the case presented here, these two
couplings are solutions of the self-duality condition (\ref{SeldDuality})
with $r=J_2/J_1=3$. Numerically, $J_1\simeq 0.4812$ and $J_2\simeq 1.4436$.
Neighboring spins linked by strong couplings $J_2$ are preferably in the same
state for inverse temperatures $\beta={1\over k_BT}\lesssim  {1\over J_2}
\simeq 0.6927$. As can be seen on Fig.~\ref{m-q_2}, a quick variation
of magnetization is indeed observed around this temperature. Weak couplings
introduce a second inverse temperature scale $\beta={1\over J_1}\simeq 2.079$
which is, in contrast, further from the second fast variation of
magnetization which occurs already around $\beta\simeq 1.50$.
\\

\begin{figure}
\centering
\ifpsfrag
\psfrag{chi}[Bc][Bc][1][1]{$\overline{\chi}$}
\psfrag{beta}[tc][tc][1][0]{$\beta$}
\psfrag{L=32}[Bc][Bc][1][0]{\tiny $L=32$}
\psfrag{L=64}[Bc][Bc][1][0]{\tiny $L=64$}
\psfrag{L=48}[Bc][Bc][1][0]{\tiny $L=48$}
\psfrag{L=96}[Bc][Bc][1][0]{\tiny $L=96$}
\setbox1=\hbox{\includegraphics[height=8.5cm]{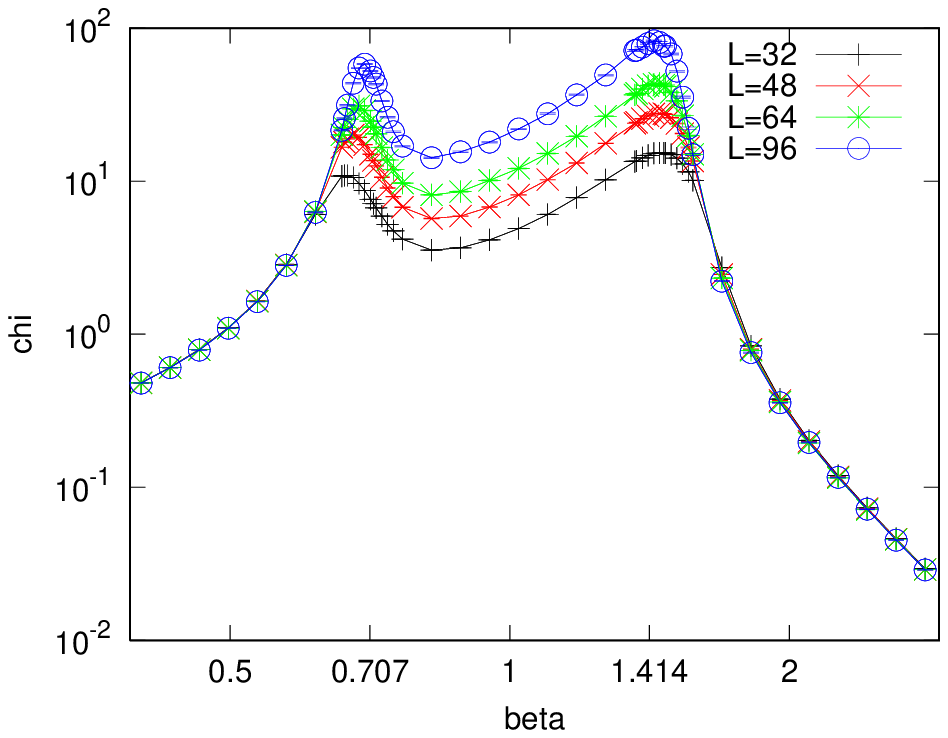}}
\includegraphics[width=8.5cm]{chi-q_2-r_3-y_0.75.eps}
\llap{\makebox[\wd1][c]{\raisebox{1cm}{\hskip 2.5cm
\includegraphics[height=2.75cm]{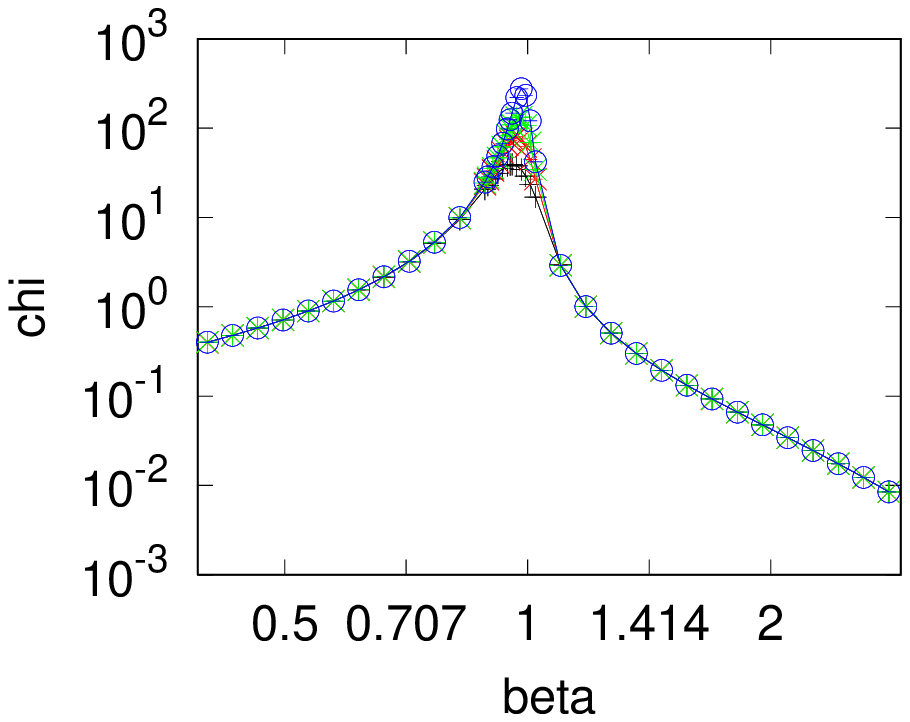}}}}
\else
\includegraphics[width=8.9cm]{Fig2.ps}
\fi
\caption{(Color online) Average susceptibility of the Ising model ($q=2$), with
a disorder strength $r=3$ and a correlation exponent $a=0.4$ ($y=0.75$), versus
the inverse temperature $\beta=1/k_BT$. The different curves correspond to
different lattice sizes ($L=32,48,64$ and 96). In the inset, the average
susceptibility in the case of uncorrelated disorder is plotted.
}\label{chi-q_2}
\end{figure}

\begin{figure}
\centering
\ifpsfrag
\psfrag{p}[Bc][Bc][1][1]{$p$}
\psfrag{chi}[Bc][Bc][1][0]{$\chi$}
\includegraphics[width=2.75cm]{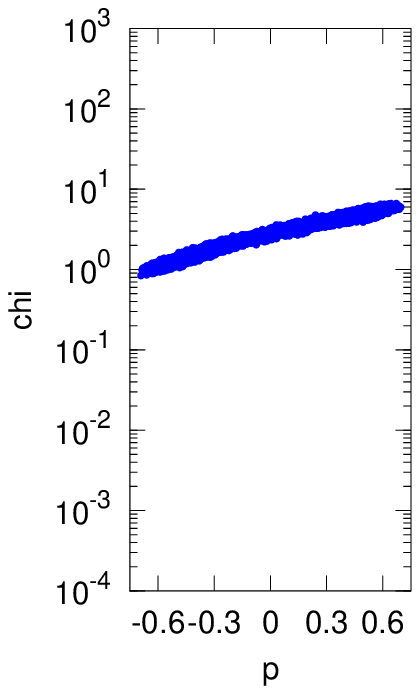}
\includegraphics[width=2.75cm]{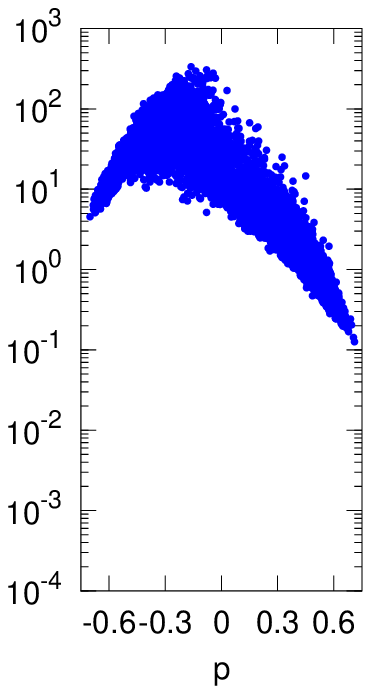}
\includegraphics[width=2.75cm]{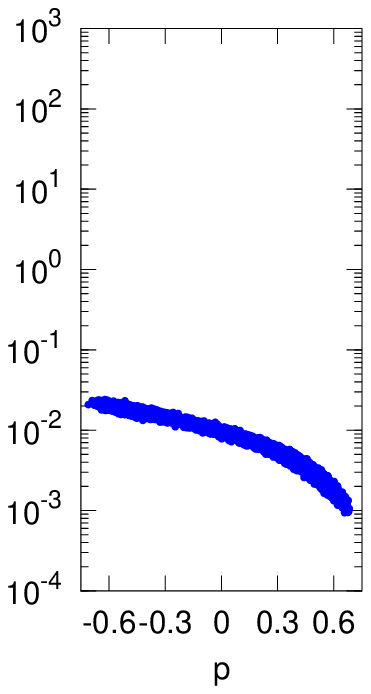}
\else
\includegraphics[width=8.9cm]{Fig3.ps}
\fi
\caption{(Color online) Susceptibility $\chi$ of the Ising model with correlated
disorder ($a=0.4$, $r=3$) versus the polarization density $p$ of the disorder
realization for $\beta\simeq 0.535$ (left), $1.022$ (center) and $2.801$
(right). Each point corresponds to a different disorder realization. The
lattice size is $L=96$.}
\label{Serie_Chi}
\end{figure}

The magnetic susceptibility, plotted in Fig.~\ref{chi-q_2}, displays two
peaks, in contrast to the single peak observed in the case of uncorrelated
disorder (see the inset of Fig.~\ref{chi-q_2}). These two peaks appear
at temperatures similar to those for which an abrupt variation of magnetization
was observed. The magnetic susceptibility diverges algebraically with the
lattice size for all temperatures in the region between the two peaks (note
the use of a logarithmic scale for the $y$-axis on the figure).
The critical exponent
associated to this divergence will be discussed in the next section.
We are therefore in presence of a Griffiths phase, like in the McCoy-Wu
model. As mentioned in the introduction, the occurrence of such a Griffiths
phase in the McCoy-Wu model was explained by the existence of exponentially
rare macroscopic regions that can order independently of the rest of the
system. The susceptibility of each disorder configuration is plotted
on Fig.~\ref{Serie_Chi} versus the polarization density $p=\langle
\sigma_i\tau_i\rangle$ of the auxiliary Ashkin-Teller model. By construction
(\ref{PolAT-Coupling}), $p$ depends linearly on the concentration of strong
couplings $J_2$. The value $p=-1$ corresponds to the configuration with only
weak couplings $J_1$ while all couplings are strong when $p=+1$. As can be
observed on Fig.~\ref{Serie_Chi}, the largest susceptibility is observed
for disorder configurations with different polarizations as the temperature
is increased. In the paramagnetic phase (left of Fig.~\ref{Serie_Chi}),
the largest susceptibility is due to configurations with a high concentration
of strong couplings. To understand this behavior, consider the two
configurations with identical couplings, either $J_1$ or $J_2$. The
system is expected to behave like a pure Potts model. Therefore, at
large temperature $\beta J_1<\beta J_2<\beta_c$, a lattice of strong
couplings $J_2$ is closer to the transition point than a configuration
with mainly weak couplings $J_1$. Analogously, in the
ferromagnetic phase (right of Fig.~\ref{Serie_Chi}), the average
susceptibility is dominated by disorder configurations with small
concentrations of strong couplings. In the Griffiths phase (center of Fig.
\ref{Serie_Chi}), the main contribution comes from disorder configurations
with a slightly negative polarization, i.e. a number of strong couplings
slightly below the number of weak ones. In such configurations, different
clusters of weak or strong couplings are typically observed. The average
magnetization is around $0.4$, which means that the largest cluster
of strong couplings occupies at most $40\%$ of the sites. There is therefore
plenty of space left for other clusters, possibly macroscopic too. Note that
the width of the bunch of points in Fig.~\ref{Serie_Chi} increases in the
Griffiths phase: at fixed polarization, some disorder configurations lead
to susceptibilities $\sim 60$ times larger than others. In the case of
uncorrelated disorder, we did not find any temperature, among those studied,
for which the plot of susceptibilities versus polarization looks like
Fig.~\ref{Serie_Chi} in the Griffiths phase.
\\

\begin{figure}
\centering
\ifpsfrag
\psfrag{Rm}[Bc][Bc][1][1]{$\overline{\langle m\rangle^2}
\ \!/\ \!\overline{\langle m\rangle}^2-1$}
\psfrag{beta}[tc][tc][1][0]{$\beta$}
\psfrag{L=32}[Bc][Bc][1][0]{\tiny $L=32$}
\psfrag{L=64}[Bc][Bc][1][0]{\tiny $L=64$}
\psfrag{L=48}[Bc][Bc][1][0]{\tiny $L=48$}
\psfrag{L=96}[Bc][Bc][1][0]{\tiny $L=96$}
\setbox1=\hbox{\includegraphics[height=8.5cm]{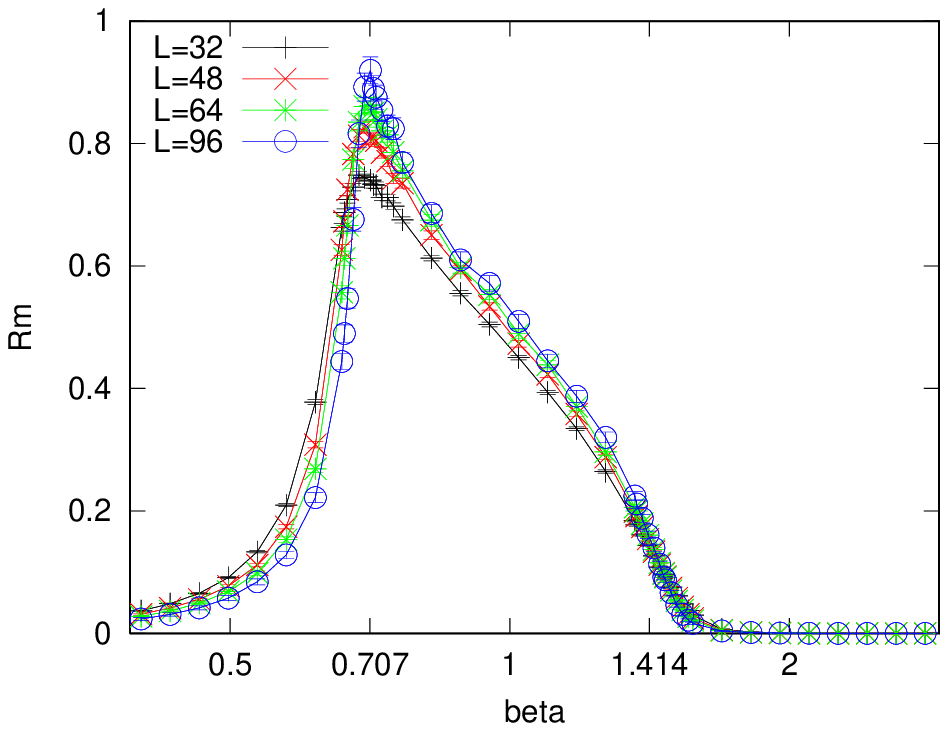}}
\includegraphics[width=8.5cm]{Rm-q_2-r_3-y_0.75.eps}
\llap{\makebox[\wd1][r]{\raisebox{3.cm}{
\includegraphics[height=2.75cm]{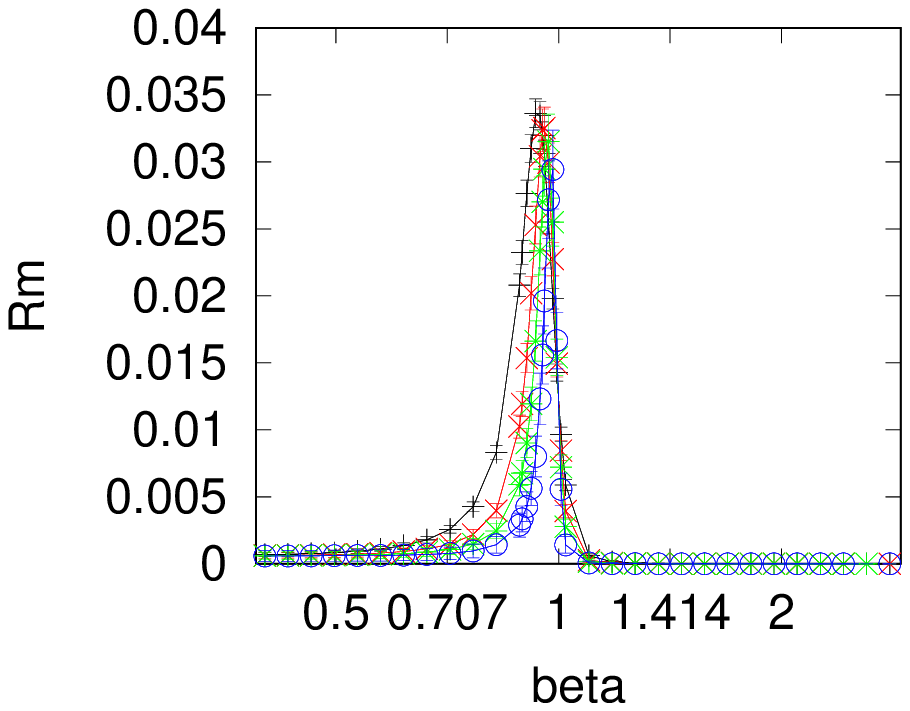}}}}
\else
\includegraphics[width=8.9cm]{Fig4.ps}
\fi
\caption{(Color online) Self-averaging ratio $R_m$ of magnetization of the
Ising model ($q=2$), with a disorder strength $r=3$ and a correlation exponent
$a=0.4$ ($y=0.75$), versus the inverse temperature $\beta=1/k_BT$. In the inset,
ratio in the case of uncorrelated disorder.
}\label{Rm-q_2}
\end{figure}

In the case of uncorrelated disorder, it is well known that sample-to-sample
fluctuations increase drastically with the lattice size at the critical point.
One consequence is the lack of self-averaging of thermodynamic quantities
when randomness is a relevant perturbation.
Averages are then dominated by rare, rather than typical, events.
In the (uncorrelated) random Potts model, self-averaging is broken only
at the critical point. Below and above, self-averaging is restored in the
thermodynamic limit. In the correlated case, magnetization is a non
self-averaging quantity in the whole Griffiths phase. Following Wiseman
and Domany~\cite{Wiseman}, the lack of self-averaging of magnetization is
measured by comparing the variance with the average:
    \be R_m={\overline{\langle m\rangle^2}-\overline{\langle m\rangle}^2
      \over\overline{\langle m\rangle}^2}.\label{Rm}\ee
This ratio is expected to go to a non-vanishing limit in the thermodynamic
limit when self-averaging is not satisfied. As can be seen on
Fig.~\ref{Rm-q_2}, this is the case in the Griffiths phase. The value
taken by $R_m$ in the thermodynamic limit is expected to be
universal~\cite{Aharony}. While no dependence on the strength of disorder
is observed, $R_m$ is however not constant in the Griffiths phase but
depends on the temperature. This implies that the critical behavior in the
Griffiths phase is not described by a single universality class. We will
come back to that point in next section. The case of the Ising model
with uncorrelated disorder is very different: as shown in the inset of
Fig.~\ref{Rm-q_2}, the ratio $R_m$ displays a thiner and thiner peak
whose maximum is close to the critical temperature. The maximum of
this peak is expected to vanish in the thermodynamic limit since
randomness is irrelevant. A similar ratio $R_\chi$ for the susceptibility
was indeed shown to decrease slowly as $1/\ln L$~\cite{Hasenbusch}.
Such a slow decay is also observed in the inset of Fig.~\ref{Rm-q_2},
though the maxima are still compatible within error bars.
\\

\begin{figure}
\centering
\ifpsfrag
\psfrag{C}[Bc][Bc][1][1]{$\overline{C}$}
\psfrag{beta}[tc][tc][1][0]{$\beta$}
\psfrag{L=32}[Bc][Bc][1][0]{\tiny $L=32$}
\psfrag{L=64}[Bc][Bc][1][0]{\tiny $L=64$}
\psfrag{L=48}[Bc][Bc][1][0]{\tiny $L=48$}
\psfrag{L=96}[Bc][Bc][1][0]{\tiny $L=96$}
\setbox1=\hbox{\includegraphics[height=8.5cm]{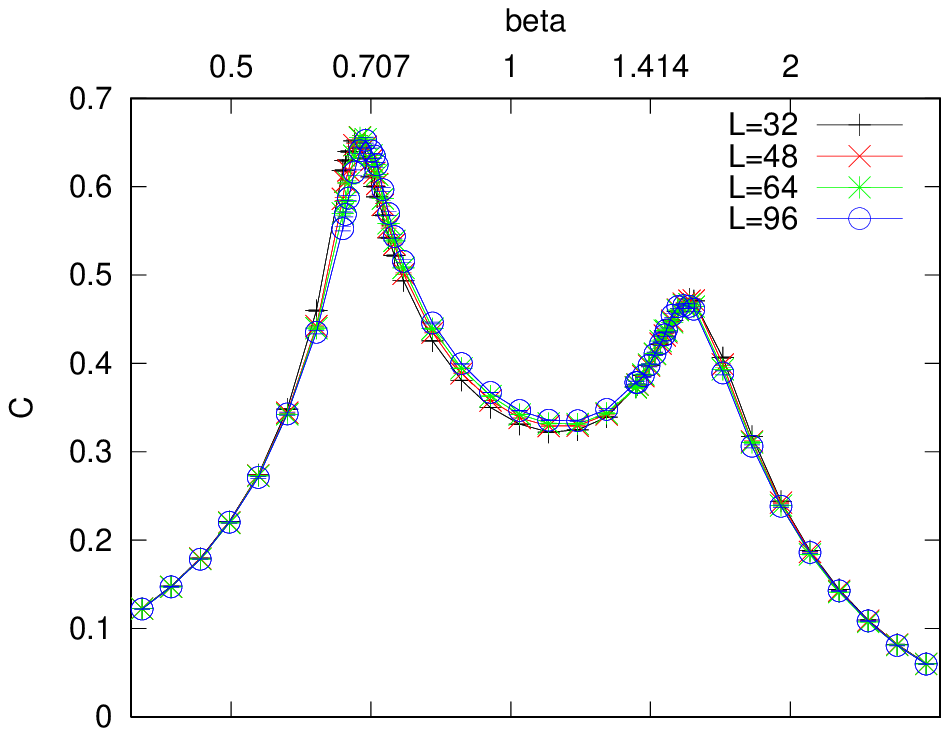}}
\includegraphics[width=8.5cm]{C-q_2-r_3-y_0.75.eps}
\llap{\makebox[\wd1][c]{\hskip 3cm\raisebox{-0.5cm}
{\includegraphics[height=2.75cm]{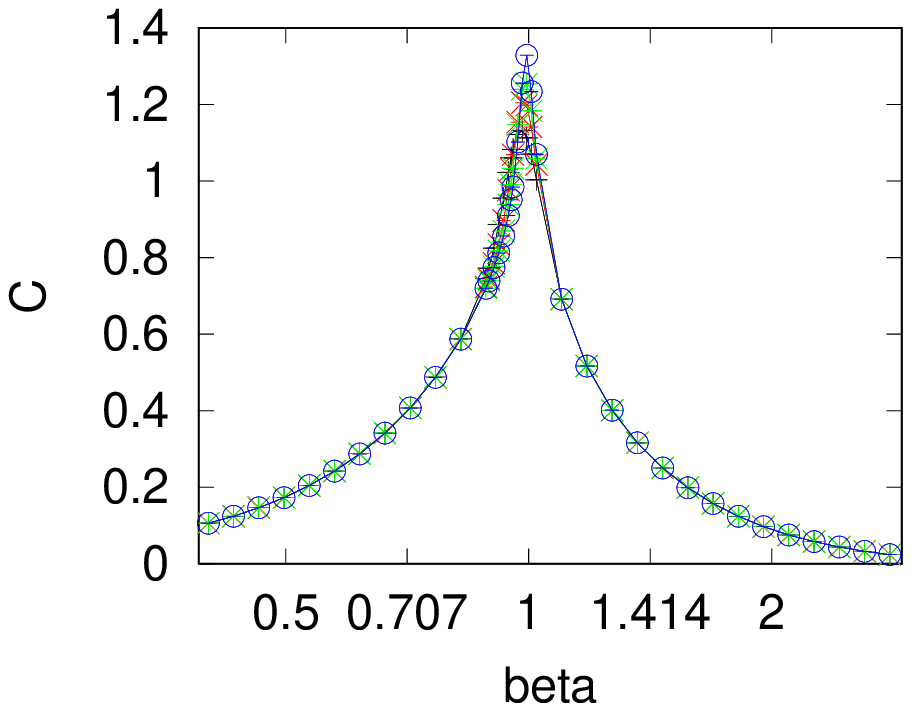}}}}
\else
\includegraphics[width=8.9cm]{Fig5.ps}
\fi
\caption{(Color online) Average specific heat of the Ising model ($q=2$), with
a disorder strength $r=3$ and a correlation exponent $a=0.4$ ($y=0.75$), versus
the inverse temperature $\beta=1/k_BT$. The different curves correspond to
different lattice sizes ($L=32,48,64$ and 96). In the inset, the specific
heat in the case of uncorrelated disorder is plotted.
}\label{C-q_2}
\end{figure}

Like the magnetic susceptibility, the specific heat displays two peaks
(Fig.~\ref{C-q_2}). However, it depends only weakly on the lattice size.
At the peaks, the data are compatible for all lattice sizes. In between, a
small increase of the specific heat with the lattice size is observed. Note
that the scale of the $y$-axis is only linear and not logarithmic, so the
evolution with the lattice size is extremely slow. When plotted versus
the polarization, the specific heat of the different disorder configurations
presents a much more weakly bended shape than the magnetic susceptibility.
The largest and smallest specific heats differ at most by a factor $\sim 2$.
The computation of the ratio
     \be R_e={\overline{\langle e\rangle^2}-\overline{\langle e\rangle}^2
      \over\overline{\langle e\rangle}^2}\label{Re}\ee
reveals that energy is a self-averaging quantity at all temperatures,
including the critical temperature. This is also the case for uncorrelated
disorder. To conclude, no evidence of Griffiths phase is found with thermal
quantities, in contrast to the magnetic sector.
\\

Finally, we note that the autocorrelation time $\tau$ also displays
two peaks, whereas only one peak is observed in the absence of disorder
correlation. These two peaks are found at the same locations as for
the magnetic susceptibility or specific heat. The autocorrelation time
evolves slowly with the lattice size, which means that the autocorrelation
exponent $z$ is close to zero.

\subsection{$q=8$ Potts model}
\begin{figure}
\centering
\ifpsfrag
\psfrag{<m>}[Bc][Bc][1][1]{$\overline{\langle m\rangle}$}
\psfrag{beta}[tc][tc][1][0]{$\beta$}
\psfrag{L=32}[Bc][Bc][1][0]{\tiny $L=32$}
\psfrag{L=64}[Bc][Bc][1][0]{\tiny $L=64$}
\psfrag{L=48}[Bc][Bc][1][0]{\tiny $L=48$}
\psfrag{L=96}[Bc][Bc][1][0]{\tiny $L=96$}
\setbox1=\hbox{\includegraphics[height=8.5cm]{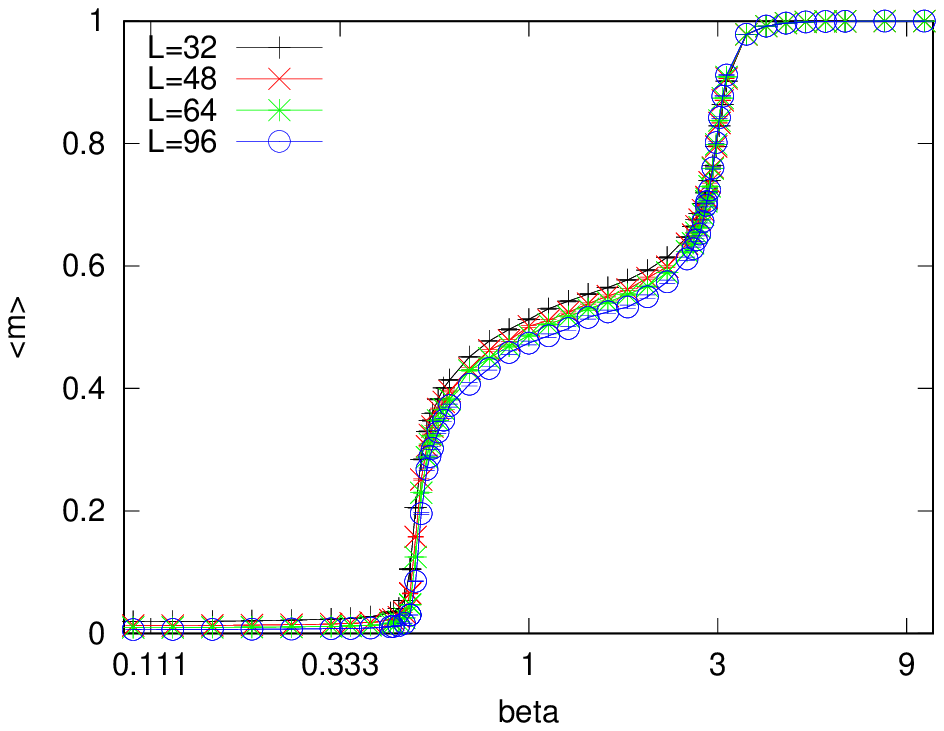}}
\includegraphics[width=8.5cm]{m-q_8-r_7.5-y_0.75.eps}
\llap{\makebox[\wd1][r]{\raisebox{0.75cm}
{\includegraphics[height=2.75cm]{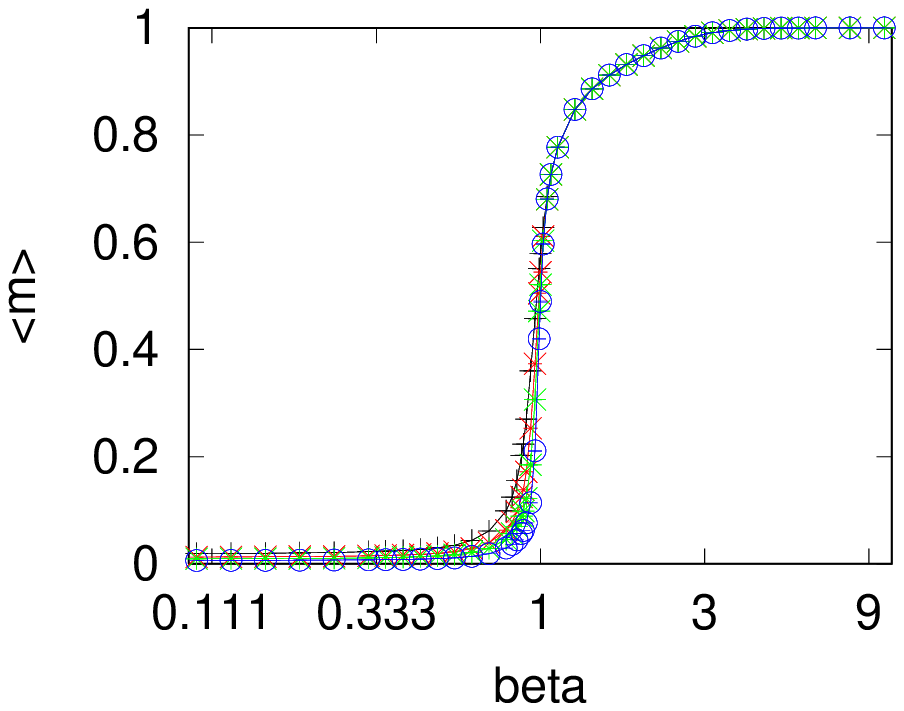}}}}
\else
\includegraphics[width=8.9cm]{Fig6.ps}
\fi
\caption{(Color online) Average magnetization of the 8-state Potts model, with
a disorder strength $r=7.5$ and a correlation exponent $a=0.4$ ($y=0.75$), versus
the inverse temperature $\beta=1/k_BT$. In the inset, magnetization curve in
the case of uncorrelated disorder.
}\label{m-q_8}
\end{figure}

\begin{figure}
\centering
\ifpsfrag
\psfrag{chi}[Bc][Bc][1][1]{$\overline{\chi}$}
\psfrag{beta}[tc][tc][1][0]{$\beta$}
\psfrag{L=32}[Bc][Bc][1][0]{\tiny $L=32$}
\psfrag{L=64}[Bc][Bc][1][0]{\tiny $L=64$}
\psfrag{L=48}[Bc][Bc][1][0]{\tiny $L=48$}
\psfrag{L=96}[Bc][Bc][1][0]{\tiny $L=96$}
\setbox1=\hbox{\includegraphics[height=8.5cm]{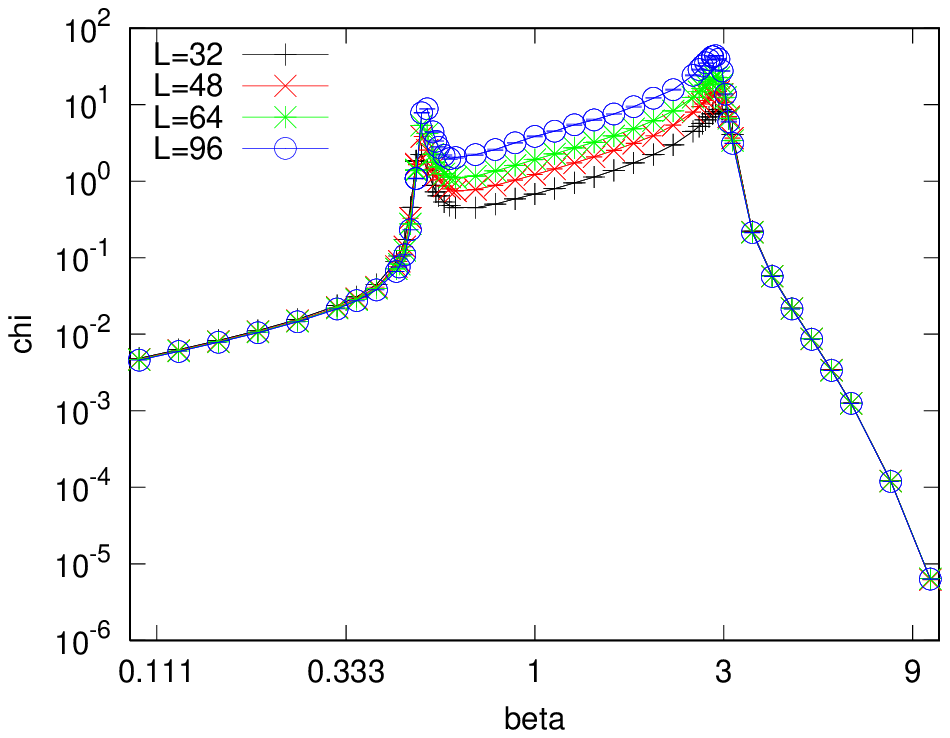}}
\includegraphics[width=8.5cm]{chi-q_8-r_7.5-y_0.75.eps}
\llap{\makebox[\wd1][c]{\raisebox{0.85cm}
{\hskip 3.5cm\includegraphics[height=2.75cm]{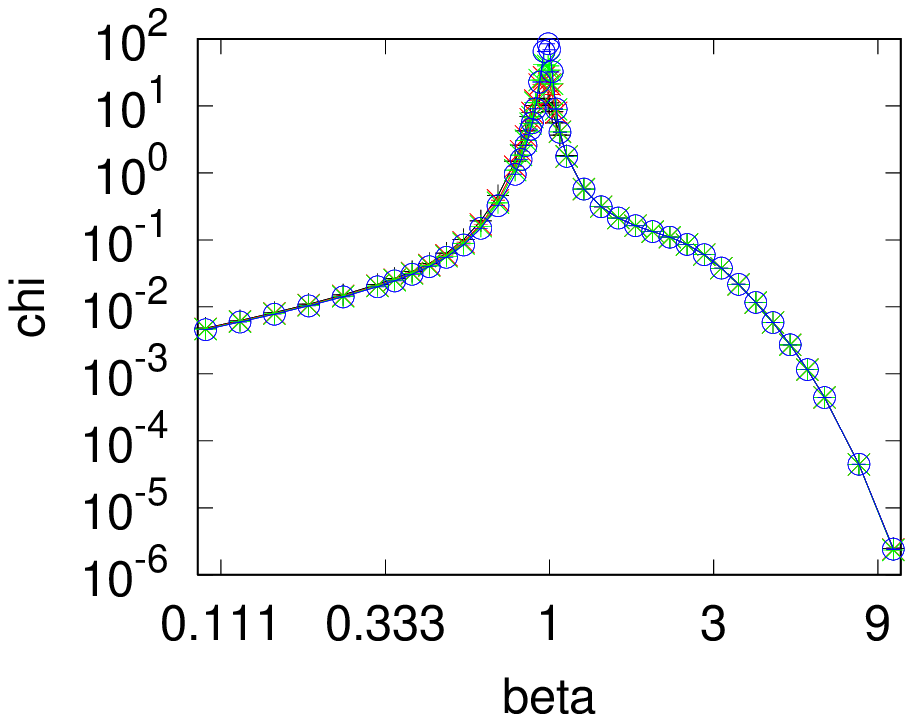}}}}
\else
\includegraphics[width=8.9cm]{Fig7.ps}
\fi
\caption{(Color online) Average susceptibility of the 8-state Potts model,
with a disorder strength $r=7.5$ and a correlation exponent $a=0.4$ ($y=0.75$),
versus the inverse temperature $\beta=1/k_BT$. In the inset, susceptibility in the
case of uncorrelated disorder.
}\label{chi-q_8}
\end{figure}

\begin{figure}
\centering
\ifpsfrag
\psfrag{Rm}[Bc][Bc][1][1]{$\overline{\langle m\rangle^2}
\ \!/\ \!\overline{\langle m\rangle}^2-1$}
\psfrag{beta}[tc][tc][1][0]{$\beta$}
\psfrag{L=32}[Bc][Bc][1][0]{\tiny $L=32$}
\psfrag{L=64}[Bc][Bc][1][0]{\tiny $L=64$}
\psfrag{L=48}[Bc][Bc][1][0]{\tiny $L=48$}
\psfrag{L=96}[Bc][Bc][1][0]{\tiny $L=96$}
\setbox1=\hbox{\includegraphics[height=8.5cm]{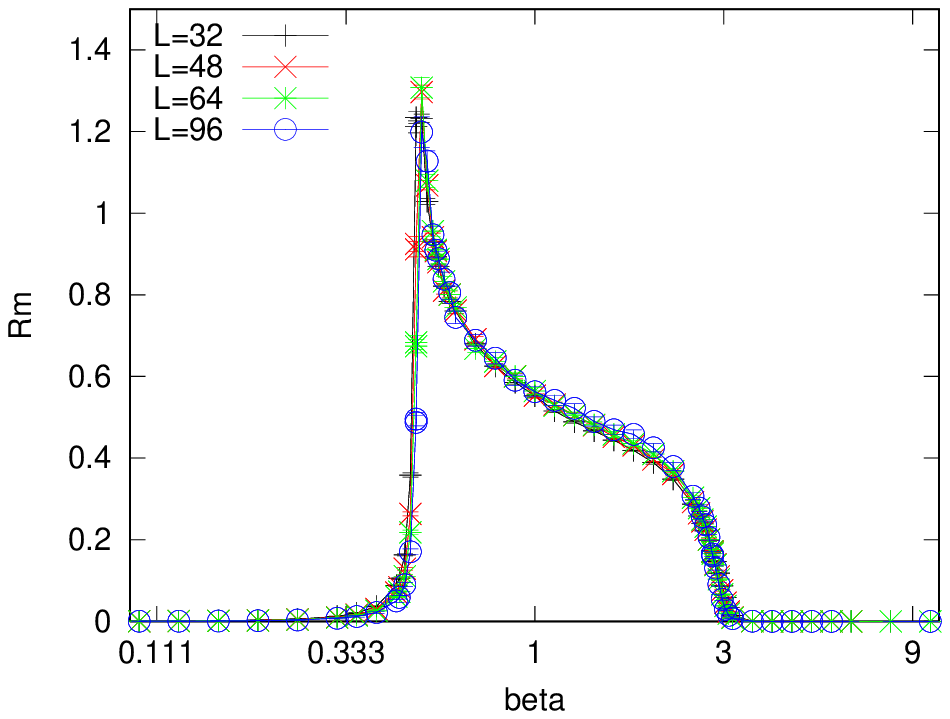}}
\includegraphics[width=8.5cm]{Rm-q_8-r_7.5-y_0.75.eps}
\llap{\makebox[\wd1][r]{\raisebox{3cm}
{\includegraphics[height=2.75cm]{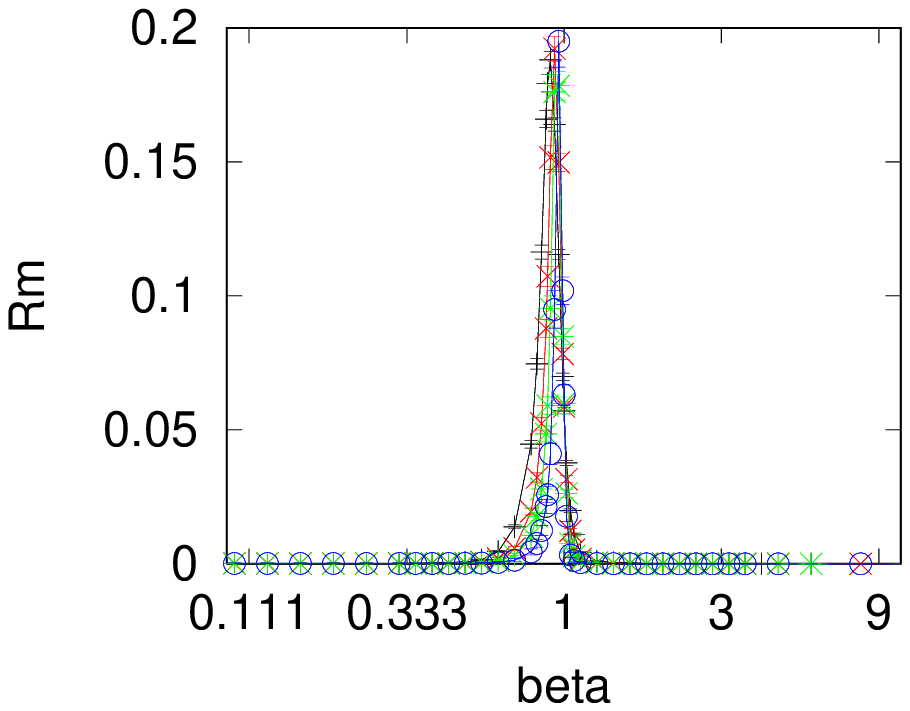}}}}
\else
\includegraphics[width=8.9cm]{Fig8.ps}
\fi
\caption{(Color online) Self-averaging ratio $R_m$ of magnetization of the
8-state Potts
model, with a disorder strength $r=7.5$ and a correlation exponent $a=0.4$
($y=0.75$), versus the inverse temperature $\beta=1/k_BT$. In the inset,
ratio in the case of uncorrelated disorder.
}\label{Rm-q_8}
\end{figure}

The 8-state Potts model with correlated disorder displays a behavior very
similar to that of the Ising model presented above. Like in the case of
uncorrelated disorder, the fact that the pure system undergoes a first-order
phase transition does not lead to any significant difference. The magnetization
curve displays two abrupt variations (Fig.~\ref{m-q_8}). The location of
these fast variations coincides with the two peaks of the magnetic
susceptibility (Fig.~\ref{chi-q_8}). For the 8-state Potts model, a stronger
disorder $r=7.5$ was considered, which means that $J_1\simeq  0.3855$
and $J_2\simeq 2.891$. In contrast to the Ising case, the first temperature
scale $\beta={1\over k_BT}={1\over J_2}\simeq 0.3459$ is significantly
smaller than the location of the first peak (around $0.50$). The second
temperature scale ${1\over J_1} \simeq 2.594$ is smaller than the location
of the second peak. When plotted versus the polarization of the auxiliary
Ashkin-Teller model, the susceptibilities of the different disorder
configurations are very similar to those of Fig.~\ref{Serie_Chi}. In the
Griffiths phase, the largest susceptibilities are again observed at a small
negative polarization ($\simeq -0.05$,) i.e. for disorder configurations
with a slightly smaller number of strong couplings than weak ones. The main
difference with the Ising model is a much larger spreading of the bunch of
points. The ratio between the largest and the smallest susceptibilities at
fixed polarization is $\sim 330$. The ratio Eq.~\ref{Rm} is plotted
on Fig.~\ref{Rm-q_8} in the case of the 8-state Potts model. Like in
the Ising case, magnetization is not self-averaging in the Griffiths phase.
\\

\begin{figure}
\centering
\ifpsfrag
\psfrag{C}[Bc][Bc][1][1]{$\overline{C}$}
\psfrag{beta}[tc][tc][1][0]{$\beta$}
\psfrag{L=32}[Bc][Bc][1][0]{\tiny $L=32$}
\psfrag{L=64}[Bc][Bc][1][0]{\tiny $L=64$}
\psfrag{L=48}[Bc][Bc][1][0]{\tiny $L=48$}
\psfrag{L=96}[Bc][Bc][1][0]{\tiny $L=96$}
\setbox1=\hbox{\includegraphics[height=8.5cm]{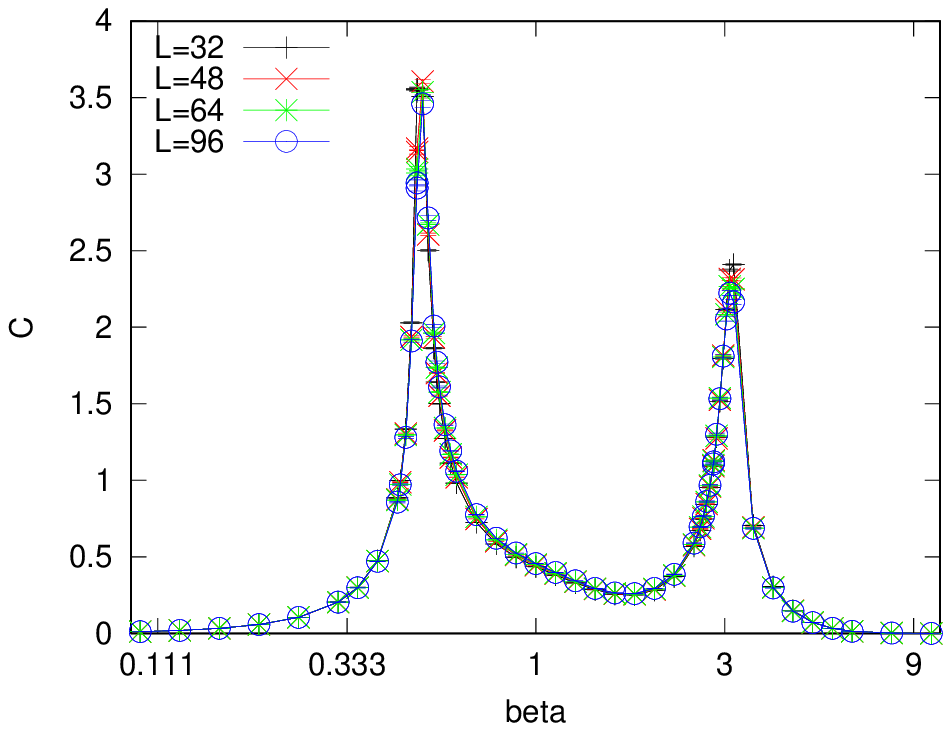}}
\includegraphics[width=8.5cm]{C-q_8-r_7.5-y_0.75.eps}
\llap{\makebox[\wd1][c]{\raisebox{4cm}
{\hskip 5.8cm\includegraphics[height=2.75cm]{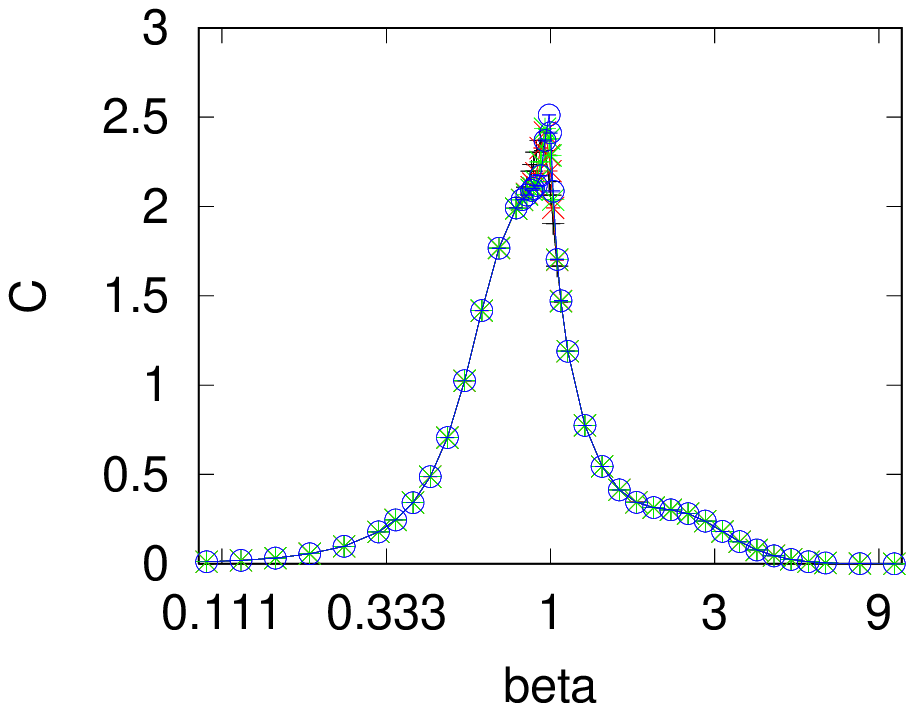}}}}
\else
\includegraphics[width=8.9cm]{Fig9.ps}
\fi
\caption{(Color online) Average specific heat of the 8-state Potts model,
with a disorder strength $r=7.5$ and a correlation exponent $a=0.4$
($y=0.75$), versus the inverse temperature $\beta=1/k_BT$. In the inset,
the specific heat in the case of uncorrelated disorder.
}\label{C-q_8}
\end{figure}

The specific heat does not display such properties. Even though two peaks
can be observed (Fig.~\ref{C-q_8}), the specific heat is essentially
size-independent, even at and between the two peaks. The ratio Eq.~\ref{Re}
vanishes at all temperatures which implies that energy is a self-averaging
quantity.
\\

Like in the Ising case, the autocorrelation time also displays two peaks
located at the same temperatures as for the magnetic susceptibility or
the specific heat. The peaks are more pronounced than in the Ising case
and the autocorrelation time reaches a maximal value $\tau\simeq 7$ at
the second peak. Again, the autocorrelation time evolves slowly with the
lattice size.

\section{\label{Sec4}Critical behavior in the Griffiths phase}
In Ref.~\cite{EPL}, the critical exponents have been estimated at the
self-dual point $\beta_c=1$ of the 8-state Potts model with correlated
disorder. In the following, the study is extended to the $q=2,4$
and $16$-state Potts models, and to several other temperatures in the
Griffiths phase. Numerical evidence of the stability of the
critical exponents against a variation of the strength of disorder
is provided.
\\

Monte Carlo simulations were performed for lattice sizes $L=16$, 24, 32, 48,
64, 96, 128, 192, and 256. The thermodynamic quantities were averaged over
a number of disorder configurations proportional to $1/L^2$. For the largest
lattice size ($L=256$), 10240 disorder configurations were generated while for
$L=64$ for example, this number is raised up to $163840$. For each disorder
configuration, 5000 Monte Carlo steps are performed.
The critical exponents $\beta/\nu$, $\gamma/\nu$ and $\nu$ are determined
by Finite-Size Scaling of the average quantities:
  \ba &&\overline{\langle m^n\rangle}^{1/n}\sim L^{-\beta/\nu},\nonumber\\
  &&\overline{\chi}=\beta L^d\big[\overline{\langle m^2\rangle}
    -\overline{\langle m\rangle^2}\big]\sim L^{\gamma/\nu},\nonumber\\
  &&-{d\ln\overline{\langle m\rangle}\over d\beta}
  =L^d{\overline{\langle me\rangle}-\overline{\langle m\rangle\langle e\rangle}
     \over \overline{\langle m\rangle}}\sim L^{1/\nu}\ea
where the moments of order $n=1,2,3$ and 4 of the magnetization are considered.
To take into account the possibility of scaling corrections, relatively strong
for the average susceptibility, power-law fits are successively performed 
in the windows $[L_{\rm min},256]$ where the smallest lattice size $L_{\rm min.}$
is iteratively increased. The different fits lead to $L_{\rm min.}$-dependent
effective critical exponents. Because of scaling
corrections, these exponents do not reach a plateau at large $L_{\rm min.}$ but
rather vary continuously with $L_{\rm min.}$. An extrapolation of these effective
exponents in the limit $1/L_{\rm min.}\rightarrow 0$ is performed. For most
of the observables, it is sufficient to consider an extrapolation with
a polynomial of degree 1 in $1/L_{\rm min.}$. 
\\

Note that the critical exponents $\beta/\nu$, $\gamma/\nu$ and $\nu$
were defined only through Finite-Size Scaling of thermodynamic
quantities. It is not clear whether a scaling law with a reduced
temperature exists or not in the thermodynamic limit. The average magnetic
susceptibility $\bar\chi$ for example diverges at all temperatures in
the Griffiths phase. It is therefore difficult to imagine how a scaling
law such as $|T-T_c|^\gamma$ could be defined for a temperature $T$
inside the Griffiths phase.

\begin{figure}
\centering
\ifpsfrag
\psfrag{m2}[Bc][Bc][1][1]{${\overline{\langle m^2\rangle}}^{1/2}$}
\psfrag{L}[Bc][Bc][1][0]{$L$}
\psfrag{1/Lmin}[tc][tc][1][0]{$1/L_{\rm min.}$}
\psfrag{beta/nu}[Bc][Bc][1][0]{$-\beta/\nu$}
\psfrag{y=0.00}[Bc][Bc][1][0]{\tiny $y=0.00$}
\psfrag{y=0.25}[Bc][Bc][1][0]{\tiny $y=0.25$}
\psfrag{y=0.50}[Bc][Bc][1][0]{\tiny $y=0.50$}
\psfrag{y=0.75}[Bc][Bc][1][0]{\tiny $y=0.75$}
\psfrag{y=1.00}[Bc][Bc][1][0]{\tiny $y=1.00$}
\psfrag{y=1.25}[Bc][Bc][1][0]{\tiny $y=1.25$}
\psfrag{Uncorr.}[Bc][Bc][1][0]{\tiny{\rm Uncorr.}}
\makebox[9.25cm][r]{
\setbox1=\hbox{\includegraphics[height=8.5cm]{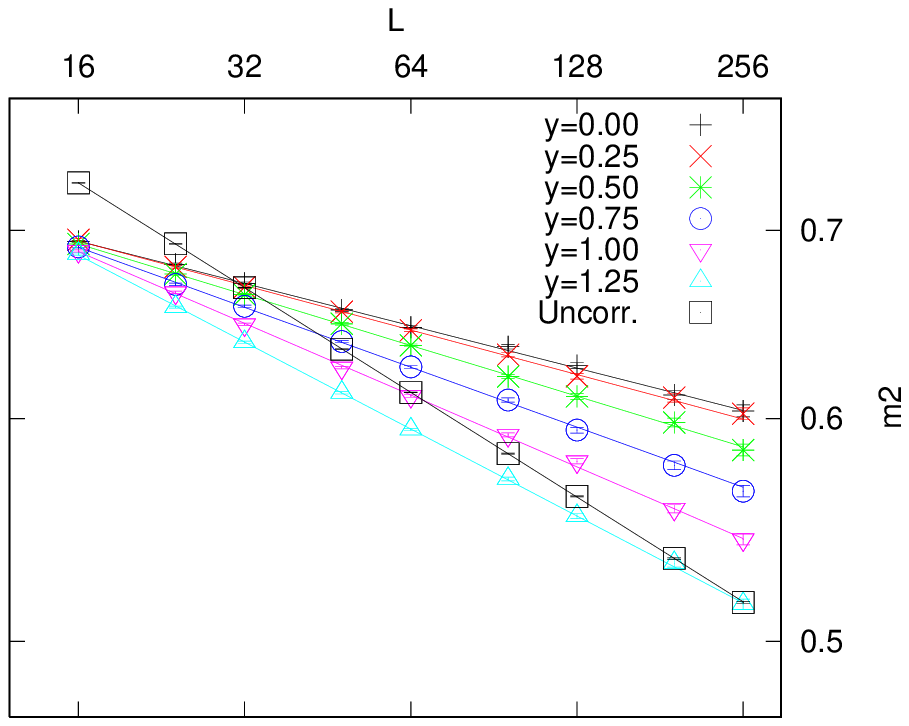}}
\includegraphics[width=8.5cm]{m-q_2-r_3-vs_L.eps}
\llap{\makebox[\wd1][l]{\raisebox{-0.5cm}
{\hskip 2cm\includegraphics[height=3.4cm]{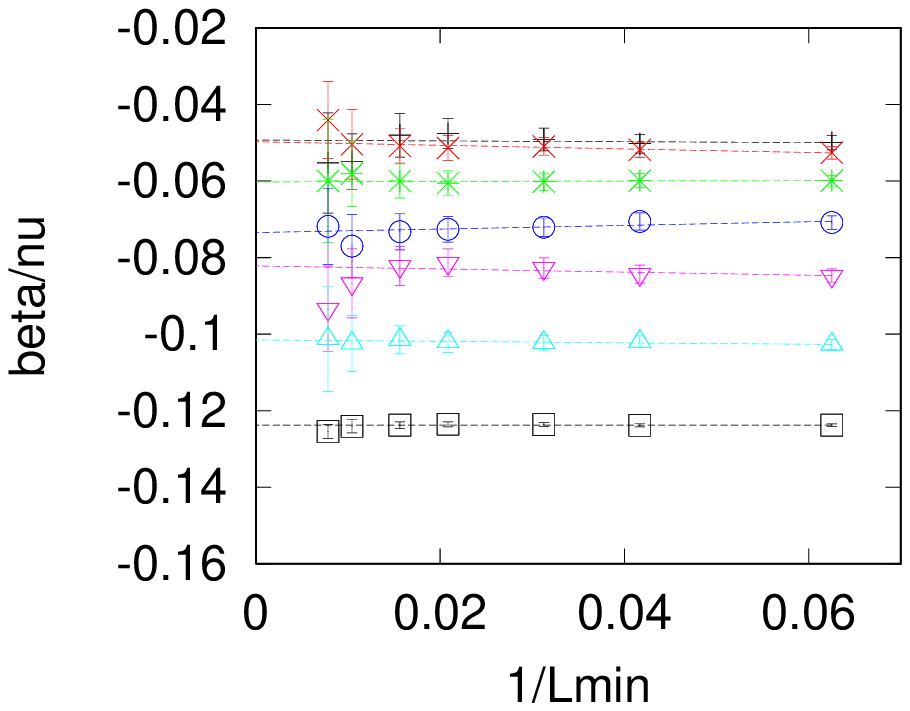}}}}}
\else
\includegraphics[width=8.9cm]{Fig10.ps}
\fi
\caption{(Color online) Finite-Size Scaling of the second moment
${\overline{\langle m^2
\rangle}}^{1/2}$ of the magnetization of the Ising model ($q=2$) with a
disorder strength $r=3$ at the self-dual point $\beta_c=1$. The different
curves correspond to different disorder correlation exponents, referred
to by the parameter $y$ of the auxiliary Ashkin-Teller model.
The curve with the legend Uncorr. corresponds to the
Ising model with uncorrelated disorder. In the inset, effective exponents
obtained by fitting the data in the window $[L_{\rm min};256]$ versus
$1/L_{\rm min}$. The straight line is a linear fit of these exponents.
}\label{m-vs-L}
\end{figure}

\begin{table*}
\caption{\label{Tab2}Critical exponent $\beta/\nu$ extrapolated from the
Finite-Size Scaling of the second moment $\overline{\langle m^2\rangle}^{1/2}$
of magnetization at the self-dual point $\beta_c=1$. The estimates for
uncorrelated disorder can be compared with the exact value $1/8$ ($q=2$)
and the transfer matrix estimates $0.1419(1)$ ($q=4$), $0.1514(2)$ ($q=8$)
from Ref.~\cite{BBCCLev}.}
\begin{ruledtabular}
\begin{tabular}{l|llllll|l}
$y$ & $0$ & $0.25$ & $0.5$ & $0.75$ & $1$ & $1.25$ & Uncorr. Dis. \\
\hline
$q=2$, $r=2$ & $0.046(6)$ & $0.051(7)$ & $0.059(6)$ & $0.069(6)$ & $0.078(5)$ & $0.102(7)$ & $0.1250(7)$\\
$q=2$, $r=3$ & $0.049(7)$ & $0.050(5)$ & $0.060(4)$ & $0.073(5)$ & $0.082(6)$ & $0.102(4)$ & $0.1238(11)$\\
\hline
$q=4$, $r=4$ & $0.052(6)$ & $0.049(7)$ & $0.064(6)$ & $0.071(6)$ & $0.088(7)$ & $0.102(5)$ & $0.139(2)$\\
\hline
$q=8$, $r=6$ & $0.053(6)$ & $0.057(7)$ & $0.066(7)$ & $0.075(6)$ & $0.092(6)$ & $0.104(5)$ & $0.150(2)$\\
$q=8$, $r=7.5$ & $0.052(6)$ & $0.059(5)$ & $0.067(6)$ & $0.076(6)$ & $0.087(5)$ & $0.104(4)$ & $0.150(2)$\\
$q=8$, $r=9$ & $0.052(5)$ & $0.056(7)$ & $0.068(7)$ & $0.074(6)$ & $0.088(5)$ & $0.108(8)$ & $0.149(3)$\\
\hline
$q=16$, $r=10$ & $0.052(5)$ & $0.056(6)$ & $0.066(7)$ & $0.072(7)$ & $0.089(5)$ & $0.108(5)$ & $0.159(3)$\\
\end{tabular}
\end{ruledtabular}
\end{table*}

Among the various moments of magnetization that were considered, the second
one displays the smallest scaling corrections. As can be seen on Fig.~%
\ref{m-vs-L} in the case of the Ising model, the effective exponent
$\beta/\nu$ does not vary much with the lowest lattice size $L_{\rm min.}$
entering into the power-law fit of $\overline{\langle m^2\rangle}^{1/2}$.
For the average magnetization and the moments of order 3 and 4, a slow
linear variation of these exponents is observed. A linear extrapolation
leads in the limit $L_{\rm min.}\rightarrow 0$ to exponents compatible
with those obtained from the second moment. The different exponents
at the self-dual point $\beta_c=1$ are collected in Table~\ref{Tab2}.
The exponents do not show any significant dependence on the strength of
disorder $r$. As already observed in the case of uncorrelated disorder,
the amplitude of the scaling corrections depends on $r$. More
interesting is the fact that the exponents $\beta/\nu$ do not depend on
the number of states $q$ of the Potts model. As mentioned in the
introduction, such a behavior is also observed in the generalization
of the McCoy-Wu model to Potts spins. However, the estimates of
$\beta/\nu$ are remarkably different from the exact value $\beta/\nu
=(3-\sqrt 5)/4$ at the McCoy-Wu-Fisher fixed point. Furthermore,
it can be observed in Table~\ref{Tab2} that the exponent $\beta/\nu$
increases when the disorder correlations decay faster, i.e. when $y$, and
therefore $a$, increases. $\beta/\nu$ remains always smaller than
in the case of uncorrelated disorder. Such a behavior was also observed
for the McCoy-Wu model with correlated disorder in the longitudinal
direction~\cite{Rieger99}. However, the magnetic scaling dimension was
shown to be $\beta/\nu\simeq a/2$ in this model while this exponent
is closer to $a/5$ in the Potts model with correlated disorder.
The dependence of $\beta/\nu$ on $a$ also contradicts Weinrib and Halperin
calculation for which $\beta/\nu={\cal O}(\epsilon^2)$ in 2D.
Therefore, the Potts model with the correlated disorder considered
in this paper belongs to a distinct universality class.
\\

The independence of the exponents with the number of states $q$
contrasts with the case of the Potts model with uncorrelated disorder
for which an increase of $\beta/\nu$ with $q$ was shown. 
A very small dependence of $\beta/\nu$ on $q$, compatible with error
bars, cannot be completely excluded. Note that in the case of the McCoy-Wu model
with correlated disorder, the exponent $\beta/\nu$ is a continuous function
of the correlation exponent $a$, even at $a=1$ corresponding in this case
to uncorrelated disorder~\cite{Rieger99}. If the same occurs for the isotropic
Potts model with correlated disorder, then one should expect a dependence
on $q$ for exponents $a<2$ because such a dependence exists for
uncorrelated disorder, i.e. for $a\ge 2$. This hypothesis could be tested
with values of $a$ close to $2$. Unfortunately, the use of the Ashkin-Teller
as an auxiliary model to generate the disorder configurations does
not allow to go beyond $a=3/4$ ($y=4/3)$ and therefore closer to the
point $a=2$. Another possible scenario is that the RG flow for the Potts model
with correlated disorder is similar to the case
studied by Weinrib and Halperin. For small values $a<d=2$, the
independence of the exponent $\beta/\nu$ on $q$ could be explained by
a critical behavior which is controlled by the same
correlated-disorder fixed point for all Potts models. Above $a=d=2$,
the latter becomes unstable and the critical behavior is then governed
by the short-range or uncorrelated fixed point where the exponents are
known to be $q$-dependent.

\begin{table*}
\caption{\label{Tab8}Critical exponent $\beta/\nu$ extrapolated from the
Finite-Size Scaling of the second moment $\overline{\langle m^2\rangle}^{1/2}$ of
magnetization at different temperatures in the Griffiths phase.}
\begin{ruledtabular}
\begin{tabular}{l|llllll}
$y$ & $0$ & $0.25$ & $0.5$ & $0.75$ & $1$ & $1.25$ \\
\hline
$q=2$, $\beta=1$ & $0.049(7)$ & $0.050(5)$ & $0.060(4)$ & $0.073(5)$ & $0.082(6)$ & $0.102(4)$ \\
$q=2$, $\beta=1.2$ & $0.040(6)$ & $0.044(5)$ & $0.048(4)$ & $0.050(4)$ & $0.063(4)$ & $0.043(7)$\\
\hline
$q=8$, $\beta=0.75$ & $0.057(7)$ & $0.062(5)$ & $0.073(5)$ & $0.092(6)$ & $0.110(7)$ & $0.138(6)$\\
$q=8$, $\beta=1$ & $0.052(6)$ & $0.059(5)$ & $0.067(6)$ & $0.076(6)$ & $0.087(5)$ & $0.104(4)$  \\
$q=8$, $\beta=1.5$ & $0.050(5)$ & $0.052(5)$ & $0.055(5)$ & $0.64(7)$ & $0.073(7)$ & $0.082(5)$ \\
$q=8$, $\beta=2$ & $0.041(6)$ & $0.045(5)$ & $0.052(6)$ & $0.056(6)$ & $0.061(4)$ & $0.052(4)$ \\
\end{tabular}
\end{ruledtabular}
\end{table*}

In the Griffiths phase, the dependence of the exponents on the temperature was
tested for the Ising and 8-state Potts models. Estimates of $\beta/\nu$ are
given in Table~\ref{Tab8}. In the Ising case, only one temperature was tested
and the corresponding estimates of the exponent $\beta/\nu$ are incompatible
with the value measured at the self-dual point for all parameters $y\ge 0.5$.
In the Potts case, the Griffiths phase is larger and three well separated
temperatures were considered. The exponent $\beta/\nu$ clearly increases
with the temperature for all values of $y$. The Griffiths phase is therefore
not described by a unique fixed point. This result corroborates the existence
of a non-constant asymptotic value of $R_m$, supposed to be universal,
in the Griffiths phase.

\begin{figure}
\centering
\ifpsfrag
\psfrag{chi}[Bc][Bc][1][1]{$\overline\chi$}
\psfrag{L}[Bc][Bc][1][0]{$L$}
\psfrag{1/Lmin}[tc][tc][1][0]{$1/L_{\rm min.}$}
\psfrag{gamma/nu}[Bc][Bc][1][0]{$\gamma/\nu$}
\psfrag{y=0.00}[Bc][Bc][1][0]{\tiny $y=0.00$}
\psfrag{y=0.25}[Bc][Bc][1][0]{\tiny $y=0.25$}
\psfrag{y=0.50}[Bc][Bc][1][0]{\tiny $y=0.50$}
\psfrag{y=0.75}[Bc][Bc][1][0]{\tiny $y=0.75$}
\psfrag{y=1.00}[Bc][Bc][1][0]{\tiny $y=1.00$}
\psfrag{y=1.25}[Bc][Bc][1][0]{\tiny $y=1.25$}
\psfrag{Uncorr.}[Bc][Bc][1][0]{\tiny{\rm Uncorr.}}
\makebox[9.25cm][r]{
\includegraphics[width=8.5cm]{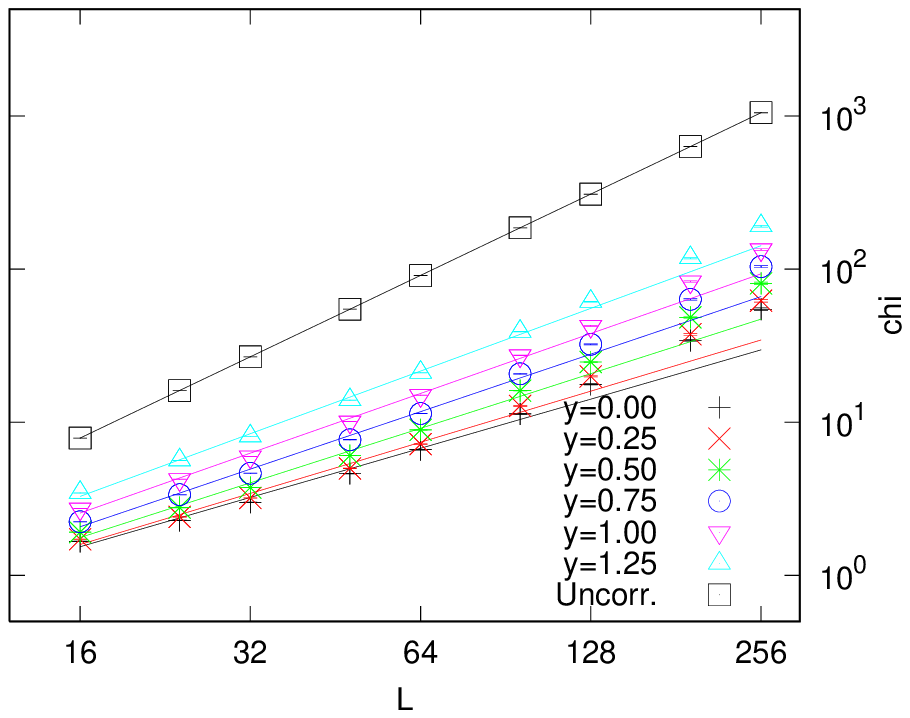}
\setbox1=\hbox{\includegraphics[height=8.5cm]{chi-q_2-r_3-vs_L.eps}}
\llap{\makebox[\wd1][l]{\raisebox{3.25cm}{\hskip 1.75cm
\includegraphics[height=3.4cm]{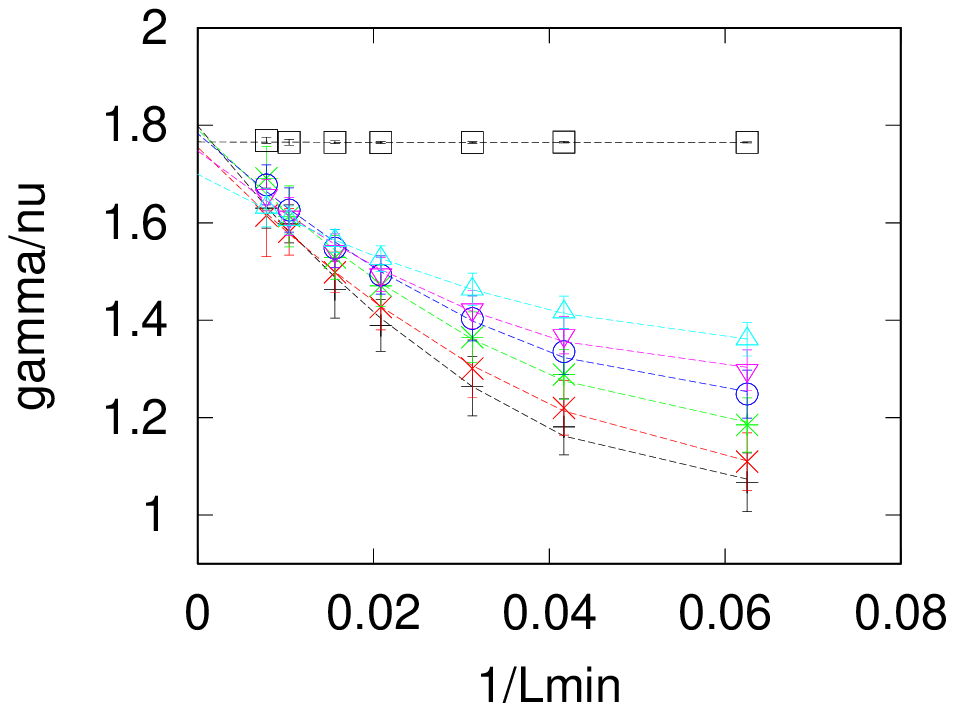}}}}}
\else
\includegraphics[width=8.9cm]{Fig11.ps}
\fi
\caption{(Color online) Finite-Size Scaling of the average susceptibility $\overline{\chi}$
of the Ising model ($q=2$) with a disorder strength $r=3$ at the critical
point $\beta_c=1$. The different curves correspond to different disorder
correlation exponents, referred to by the parameter $y$ of the auxiliary
Ashkin-Teller model. The curve with the legend Uncorr. corresponds to the
Ising model with uncorrelated disorder. In the inset, effective exponents
obtained by fitting the data in the window $[L_{\rm min};256]$ versus
$1/L_{\rm min}$. The solid line is a parabolic fit of these exponents.
}\label{chi-vs-L}
\end{figure}

\begin{table*}
\caption{(Color online) \label{Tab3b}Critical exponent $\gamma/\nu$ extrapolated from the
Finite-Size Scaling of the average susceptibility $\overline{\chi}$ at the
self-dual point $\beta_c=1$. A polynomial of degree 2 was used for the
extrapolation. The estimates for uncorrelated disorder can be compared
with the exact value $7/4$ ($q=2$) and the transfer matrix estimates
$1.7162(2)$ ($q=4$), $1.6972(4)$ ($q=8$) from Ref.~\cite{BBCCLev} assuming
that hyperscaling holds.}
\begin{ruledtabular}
\begin{tabular}{l|llllll|l}
$y$ & $0$ & $0.25$ & $0.5$ & $0.75$ & $1$ & $1.25$ & Uncorr. Dis. \\
\hline
$q=2$, $r=2$ & $1.66(7)$ & $1.73(9)$ & $1.69(12)$ & $1.66(7)$ & $1.64(6)$ & $1.64(7)$ & $1.759(6)$ \\
$q=2$, $r=3$ & $1.80(7)$ & $1.76(8)$ & $1.79(10)$ & $1.78(7)$ & $1.75(6)$ & $1.70(4)$ & $1.766(8)$ \\
\hline
$q=4$, $r=4$ & $1.84(9)$ & $1.80(7)$ & $1.80(5)$ & $1.74(5)$ & $1.81(5)$ & $1.77(7)$ & $1.717(13)$ \\
\hline
$q=8$, $r=6$ & $1.77(12)$ & $1.77(7)$ & $1.79(7)$ & $1.80(8)$ & $1.77(6)$ & $1.75(7)$ & $1.68(2)$ \\
$q=8$, $r=7.5$ & $1.77(7)$ & $1.75(8)$ & $1.78(6)$ & $1.78(8)$ & $1.75(7)$ & $1.74(5)$ & $1.70(3)$ \\
$q=8$, $r=9$ & $1.78(9)$ & $1.84(11)$ & $1.80(8)$ & $1.77(6)$ & $1.72(6)$ & $1.74(6)$ & $1.71(2)$ \\
\hline
$q=16$, $r=10$ & $1.78(10)$ & $1.80(12)$ & $1.79(9)$ & $1.78(10)$ & $1.72(6)$ & $1.74(5)$ & $1.75(3)$ \\
\end{tabular}
\end{ruledtabular}
\end{table*}

\begin{figure}
\centering
\ifpsfrag
\psfrag{chi/m}[Bc][Bc][1][1]{$\overline{\chi}\ /\ L^d\ \!\overline{\langle m\rangle}^2$}
\psfrag{L}[Bc][Bc][1][0]{$L$}
\psfrag{y=0.00}[Bc][Bc][1][0]{\tiny $y=0.00$}
\psfrag{y=0.25}[Bc][Bc][1][0]{\tiny $y=0.25$}
\psfrag{y=0.50}[Bc][Bc][1][0]{\tiny $y=0.50$}
\psfrag{y=0.75}[Bc][Bc][1][0]{\tiny $y=0.75$}
\psfrag{y=1.00}[Bc][Bc][1][0]{\tiny $y=1.00$}
\psfrag{y=1.25}[Bc][Bc][1][0]{\tiny $y=1.25$}
\psfrag{Uncorr.}[Bc][Bc][1][0]{\tiny{\rm Uncorr.}}
\includegraphics[width=8.5cm]{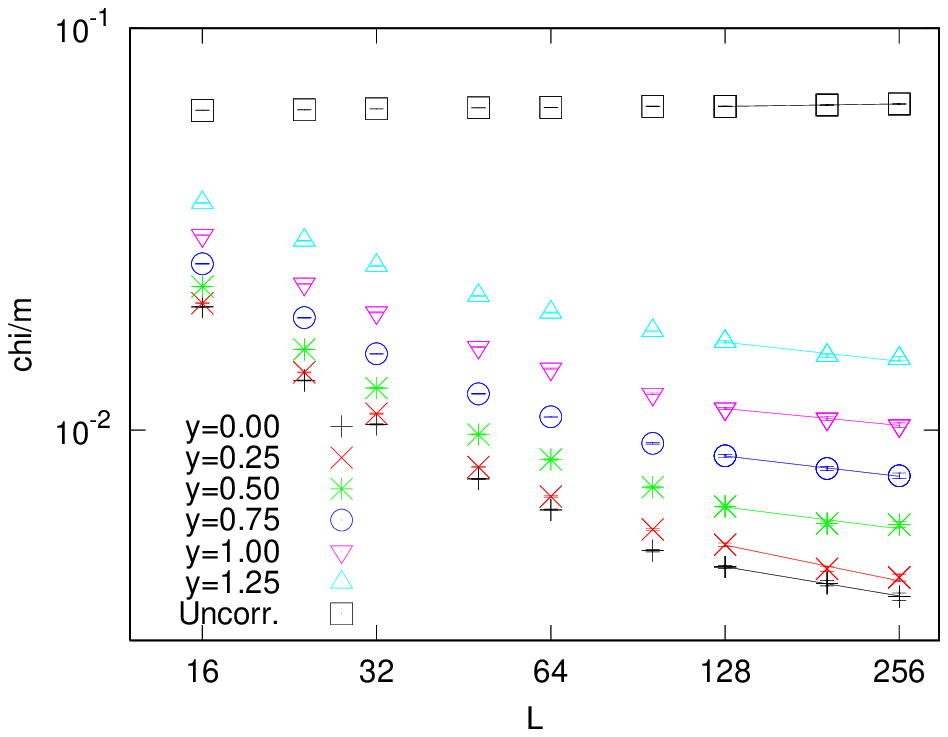}
\else
\includegraphics[width=8.9cm]{Fig12.ps}
\fi
\caption{(Color online) $\overline{\chi}\ /\ L^d\overline{\langle m\rangle}^2$ versus the
lattice size $L$ for the Ising model ($q=2$) with a disorder strength $r=3$
at the self-dual point $\beta_c=1$. According to the hyperscaling relation,
this ratio should scale as $L^0$. The straight line is a power-law fit
over the three last lattice sizes. The slopes, as given by the fit, are
$-0.24(5)$, -$0.38(10)$, $-0.3(2)$, $-0.30(5)$, $-0.31(6)$, $-0.18(3)$
for $y=0.00$, 0.25, 0.50, 0.75, 1.00 and 1.25. For uncorrelated disorder,
this value is $0.012(10)$.
}\label{Violation-vs-L}
\end{figure}

The estimation of the exponent $\gamma/\nu$ from the Finite-Size Scaling
of the magnetic susceptibility is much more difficult. Much stronger scaling
corrections are present, especially for small values of $y$. These
corrections manifest themselves on Fig.~\ref{chi-vs-L} as a gap between
the numerical data at large lattice sizes and the power-law fit
plotted as a continuous line. The effective
exponents indeed start with values in the range $[1.1;1.3]$ and then
increase as more and more small lattice sizes are removed from the fit
(see inset of Fig.~\ref{chi-vs-L}).
When only lattice sizes $L\ge 128$ are taken into account in the fit,
the effective exponents take values around $1.6-1.7$, the value reported
in Ref.~\cite{EPL}. On the figure, the effective exponents increase
with the parameter $y$ and tend towards a value close to the exponent
of the uncorrelated disorder. Because of the curvature displayed by the
effective exponents when plotted versus $1/L_{\rm min.}$, a simple linear
extrapolation, like in the case of magnetization, does not take into
account reliably the scaling corrections. It turns out that the exponents
fall nicely on the parabolic fit represented on the figure. However, the
use of a polynomial of degree 2 reduces the number of degrees of freedom of
the fit, and thus increases the error bar and lowers the stability of
the extrapolated exponents. The latter are given in Tables~\ref{Tab3b} and
\ref{Tab9b}. No significant dependence on the number of states $q$, the
strength of disorder $r$, nor the temperature in the Griffiths phase
can be noticed. While the estimates of $\gamma/\nu$ are compatible
between each other, half of them are not compatible with the
hyperscaling relation
    \be{\gamma\over\nu}=d-2{\beta\over\nu}.\label{Hyperscaling}\ee
On Fig.~\ref{Violation-vs-L}, the ratio $\overline{\chi}\ /\ L^d
\overline{\langle m\rangle}^2$ is plotted versus the lattice size
in the case of the Ising model. This quantity is expected to scale
as $L^0$ when the hyperscaling relation Eq. \ref{Hyperscaling} holds.
A power-law behavior is indeed observed at large lattice sizes,
though with a smaller exponent than at small lattice sizes. A fit over the
three largest lattice sizes leads to negative exponents, compatible
within error bars with zero, i.e. the hyperscaling relation, in
only $20\%$ of all cases considered. This statement is also true
at different temperatures in the Griffiths phase. Of course, we cannot
completely exclude the possibility of a restoration of the hyperscaling
relation at larger lattice sizes. In the case of uncorrelated disorder,
the data are in much better agreement with the hyperscaling relation.

\begin{table*}
\caption{\label{Tab9b}Critical exponent $\gamma/\nu$ extrapolated from the
Finite-Size Scaling of the average susceptibility $\overline{\chi}$ at
different temperatures in the Griffiths phase. A polynomial of degree 2 was
used for the extrapolation.
}
\begin{ruledtabular}
\begin{tabular}{l|llllll}
$y$ & $0$ & $0.25$ & $0.5$ & $0.75$ & $1$ & $1.25$ \\
\hline
$q=2$, $\beta=1$ & $1.80(7)$ & $1.76(8)$ & $1.79(10)$ & $1.78(7)$ & $1.75(6)$ & $1.70(4)$ \\
$q=2$, $\beta=1.2$ & $1.64(5)$ & $1.70(8)$ & $1.73(8)$ & $1.71(5)$ & $1.76(4)$ & $1.69(4)$ \\
\hline
$q=8$, $\beta=0.75$ & $1.80(11)$ & $1.86(8)$ & $1.78(9)$ & $1.68(6)$ & $1.71(7)$ & $1.68(4)$ \\
$q=8$, $\beta=1$ & $1.77(7)$ & $1.75(8)$ & $1.78(6)$ & $1.78(8)$ & $1.75(7)$ & $1.74(5)$ \\
$q=8$, $\beta=1.5$ & $1.80(10)$ & $1.70(8)$ & $1.72(6)$ & $1.77(6)$ & $1.78(8)$ & $1.78(5)$ \\
$q=8$, $\beta=2$ & $1.68(8)$ & $1.74(9)$ & $1.78(8)$ & $1.74(6)$ & $1.71(8)$ & $1.78(4)$ \\
\end{tabular}
\end{ruledtabular}
\end{table*}

As commented in the first section, the specific heat does not seem to display
any divergence in the Griffiths phase. Even with larger lattice sizes, up to
$L=256$, it is not possible to isolate any singular part from a regular
background. It means that the specific heat exponent $\alpha/\nu$ is therefore
smaller or equal to zero.

\begin{figure}
\centering
\ifpsfrag
\psfrag{dln/dbeta}[Bc][Bc][1][1]{$-{d\over d\beta}\ln\overline{\langle m\rangle}$}
\psfrag{L}[Bc][Bc][1][0]{$L$}
\psfrag{1/Lmin}[tc][tc][1][0]{$1/L_{\rm min.}$}
\psfrag{1/nu}[Bc][Bc][1][0]{$1/\nu$}
\psfrag{y=0.00}[Bc][Bc][1][0]{\tiny $y=0.00$}
\psfrag{y=0.25}[Bc][Bc][1][0]{\tiny $y=0.25$}
\psfrag{y=0.50}[Bc][Bc][1][0]{\tiny $y=0.50$}
\psfrag{y=0.75}[Bc][Bc][1][0]{\tiny $y=0.75$}
\psfrag{y=1.00}[Bc][Bc][1][0]{\tiny $y=1.00$}
\psfrag{y=1.25}[Bc][Bc][1][0]{\tiny $y=1.25$}
\psfrag{Uncorr.}[Bc][Bc][1][0]{\tiny{\rm Uncorr.}}
\makebox[9.25cm][r]{
\setbox1=\hbox{\includegraphics[height=8.5cm]{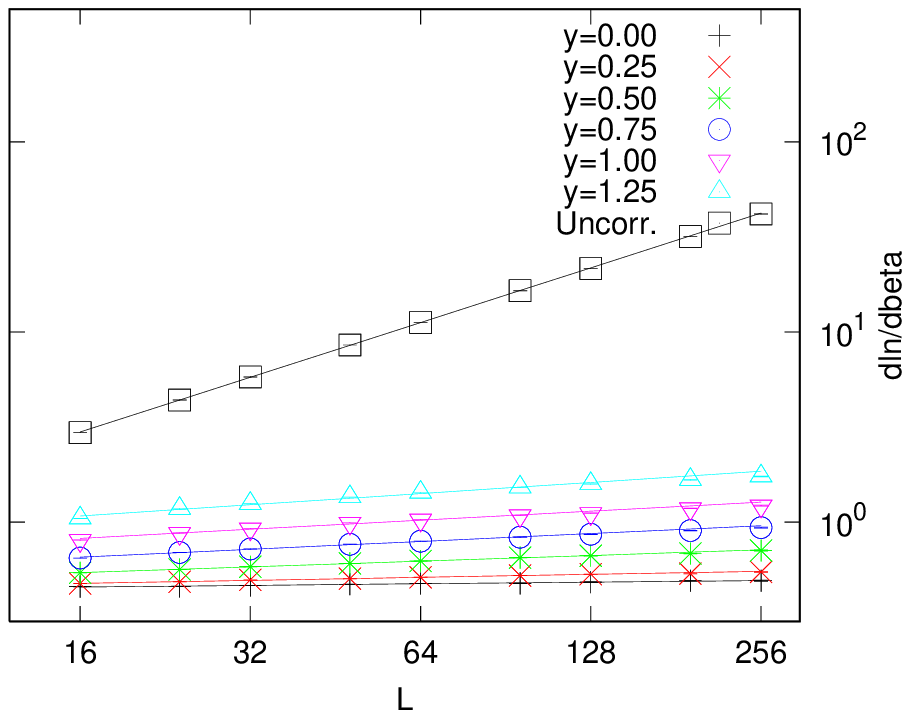}}
\includegraphics[width=8.5cm]{dln-dBeta-q_2-r_3-vs_L.eps}
\llap{\makebox[\wd1][l]{\raisebox{3.5cm}
{\hskip 2cm\includegraphics[height=3.4cm]{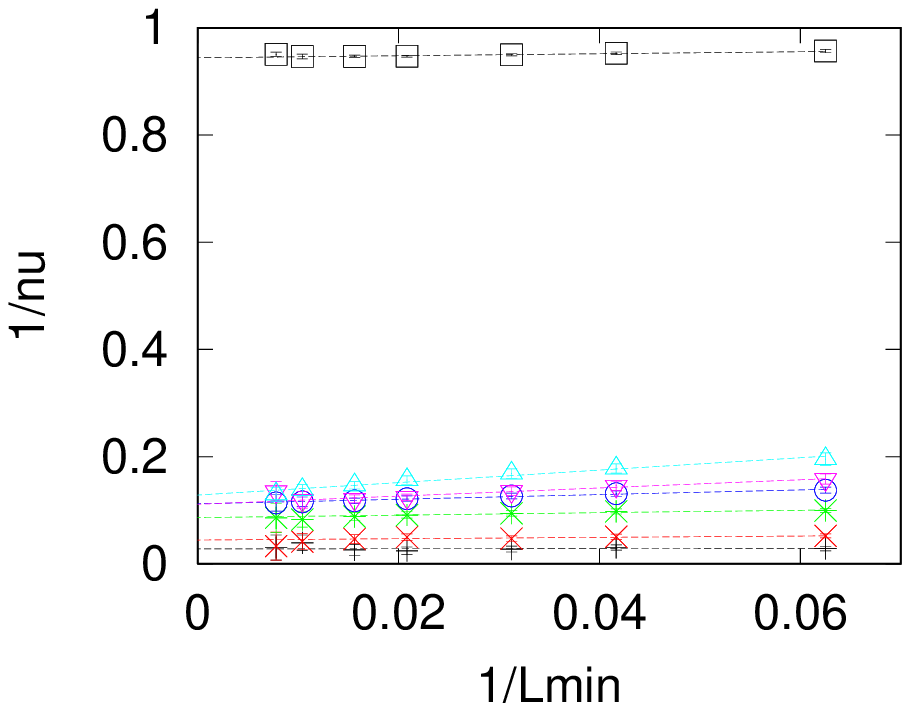}}}}}
\else
\includegraphics[width=8.9cm]{Fig13.ps}
\fi
\caption{(Color online) Finite-Size Scaling of $-{d\over d\beta}\ln\overline{\langle m\rangle}$
for the Ising model ($q=2$) with a disorder strength $r=3$ at the critical
point $\beta_c=1$. The different curves correspond to different disorder
correlation exponents, referred to by the parameter $y$ of the auxiliary
Ashkin-Teller model. The curve with the legend Uncorr. corresponds to the
Ising model with uncorrelated disorder. In the inset, effective exponents
$1/\nu$ obtained by fitting the data in the window $[L_{\rm min};256]$ versus
$1/L_{\rm min}$. The straight line is a linear fit of these exponents.
}\label{dlndBeta}
\end{figure}

The Finite-Size Scaling of the derivative of the logarithm of the average
magnetization with respect to temperature gives access to the exponent
$\nu$~\cite{Remark1}. In contrast to the magnetic susceptibility, a linear
fit of the $L_{\rm min.}$-dependent effective exponents is sufficient to
properly take into account the scaling corrections (see Fig.~\ref{dlndBeta}).
The extrapolated exponents are given in Tables~\ref{Tab6} and \ref{Tab12}.
They turn out to be much larger than in the case of uncorrelated disorder
for which $\nu\simeq 1$~\cite{NuLarge}.
In a few cases, the extrapolation procedure appears to be unstable: several
spurious values can be seen in the tables (for example for $q=8,r=7.5,y=0$
or $q=2,\beta=1.2$).
From the rest of the data, the same trends as for magnetization are observed:
the exponents do not significantly vary with the strength of disorder $r$,
nor the number of states $q$. However, in contrast to magnetization, no
significant dependence on temperature can be seen, apart for $y=1.25$.
More accurate values would be necessary. Note that the exponent $\nu$
takes values incompatible with the Weinrib-Halperin prediction $\nu=2/a$
for the $O(n)$ model with (Gaussian) correlated disorder
For comparison, the latter would give
$1/\nu=0.125$, 0.143, 0.167, 0.2, 0.25 and 0.333 for the
values of $a$ that were considered. Finally, note that the hyperscaling
relation ${\alpha\over\nu}={2\over\nu}-d$ leads, with the estimates of $\nu$
reported in Tables~\ref{Tab6} and \ref{Tab12}, to negative specific heat 
exponents ${\alpha\over\nu}<-1.7$. Because of the regular contribution to
the specific heat, this prediction cannot be tested.

\begin{table*}
\caption{\label{Tab6}Critical exponent $1/\nu$ extrapolated from the
Finite-Size Scaling of the quantity $-{d\over d\beta}\ln\overline{\langle m
\rangle}$ at the self-dual point $\beta_c=1$. For uncorrelated disorder,
estimates of $\nu$ slightly above 1, but usually compatible, were reported
in the literature~\cite{RBPM-MC,RBPotts}. In the case $q=2$, $\nu=1$ (with
logarithmic corrections) since uncorrelated disorder is marginally
irrelevant.
}
\begin{ruledtabular}
\begin{tabular}{l|llllll|l}
$y$ & $0$ & $0.25$ & $0.5$ & $0.75$ & $1$ & $1.25$ & Uncorr. Dis. \\
\hline
$q=2$, $r=2$ & $0.041(10)$ & $0.065(12)$ & $0.099(13)$ & $0.116(13)$ & $0.11(2)$ & $0.14(2)$ & $0.969(4)$\\
$q=2$, $r=3$ & $0.028(13)$ & $0.044(11)$ & $0.086(10)$ & $0.112(9)$ & $0.11(2)$ & $0.13(2)$ & $0.944(5)$\\
\hline
$q=4$, $r=4$ & $0.033(12)$ & $0.041(13)$ & $0.082(12)$ & $0.103(14)$ & $0.113(14)$ & $0.116(13)$ & $0.985(7)$\\
\hline
$q=8$, $r=6$ & $0.026(11)$ & $0.05(2)$ & $0.08(2)$ & $0.101(11)$ & $0.115(11)$ & $0.113(11)$ & $0.965(11)$\\
$q=8$, $r=7.5$ & $0.25(10)$ & $0.047(12)$ & $0.083(9)$ & $0.107(10)$ & $0.111(9)$ & $0.114(7)$ & $0.976(9)$\\
$q=8$, $r=9$ & $0.018(11)$ & $0.041(12)$ & $0.075(12)$ & $0.097(11)$ & $0.102(9)$ & $0.109(12)$ & $0.972(7)$\\
\hline
$q=16$, $r=10$ & $0.022(10)$ & $0.04(2)$ & $0.069(10)$ & $0.083(14)$ & $0.101(10)$ & $0.113(9)$ & $1.02(3)$\\
\end{tabular}
\end{ruledtabular}
\end{table*}

\begin{table*}
\caption{\label{Tab12}Critical exponent $1/\nu$ estimated from the Finite-Size
Scaling of the quantity $-{d\over d\beta}\ln\overline{\langle m\rangle}$
at different temperatures in the Griffiths phase.}
\begin{ruledtabular}
\begin{tabular}{l|llllll}
$y$ & $0$ & $0.25$ & $0.5$ & $0.75$ & $1$ & $1.25$ \\
\hline
$q=2$, $\beta=1$ & $0.028(13)$ & $0.044(11)$ & $0.086(10)$ & $0.112(9)$ & $0.11(2)$ & $0.13(2)$ \\
$q=2$, $\beta=1.2$ & $-0.03(2)$ & $0.00(2)$ & $0.05(2)$ & $0.059(12)$ & $0.13(2)$ & $0.31(2)$\\
\hline
$q=8$, $\beta=0.75$ & $0.04(2)$ & $0.06(2)$ & $0.09(2)$ & $0.11(2)$ & $0.13(2)$ & $0.187(11)$\\
$q=8$, $\beta=1$ & $0.025(10)$ & $0.047(12)$ & $0.083(9)$ & $0.107(10)$ & $0.111(9)$ & $0.114(7)$\\
$q=8$, $\beta=1.5$ & $0.03(2)$ & $0.04(2)$ & $0.07(2)$ & $0.103(14)$ & $0.10(2)$ & $0.146(13)$ \\
$q=8$, $\beta=2$ & $0.01(2)$ & $0.03(2)$ & $0.08(2)$ & $0.079(13)$ & $0.11(2)$ & $0.336(13)$ \\
\end{tabular}
\end{ruledtabular}
\end{table*}

\section{\label{Sec5}Disorder fluctuations and hyperscaling
violation}
\begin{figure}
\centering
\ifpsfrag
\psfrag{FluctM}[Bc][Bc][1][1]{$\chi_1$}
\psfrag{beta}[tc][tc][1][0]{$\beta$}
\psfrag{L=32}[Bc][Bc][1][0]{\tiny $L=32$}
\psfrag{L=64}[Bc][Bc][1][0]{\tiny $L=64$}
\psfrag{L=48}[Bc][Bc][1][0]{\tiny $L=48$}
\psfrag{L=96}[Bc][Bc][1][0]{\tiny $L=96$}
\makebox[9.25cm][r]{
\setbox1=\hbox{\includegraphics[height=8.5cm]{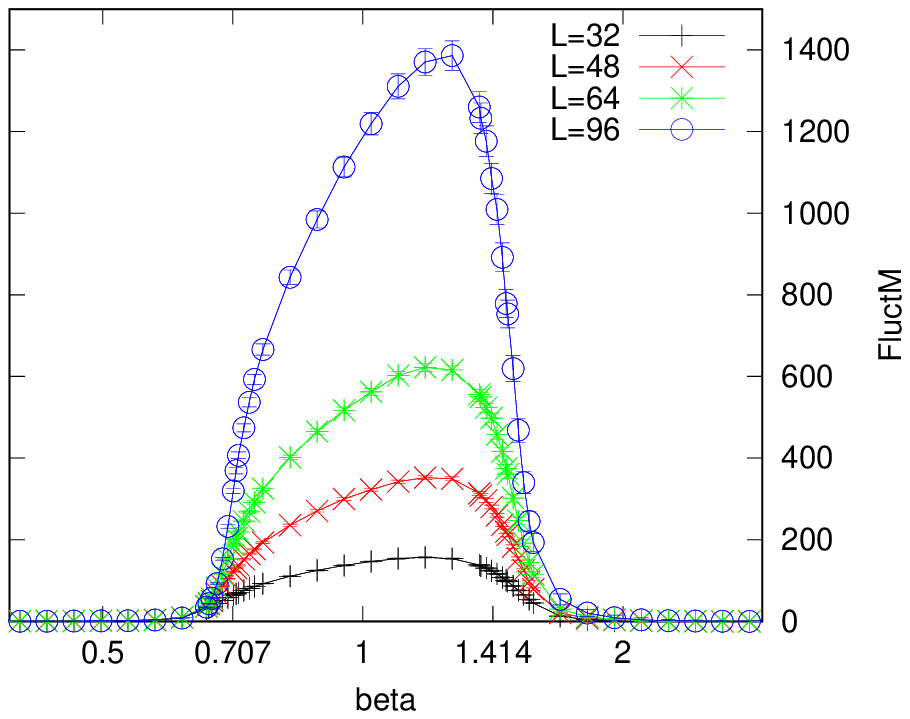}}
\includegraphics[width=8.5cm]{FluctM-q_2-r_3-y_0.75.eps}
\llap{\makebox[\wd1][l]{\raisebox{4cm}{\hskip 2cm
\includegraphics[height=2.75cm]{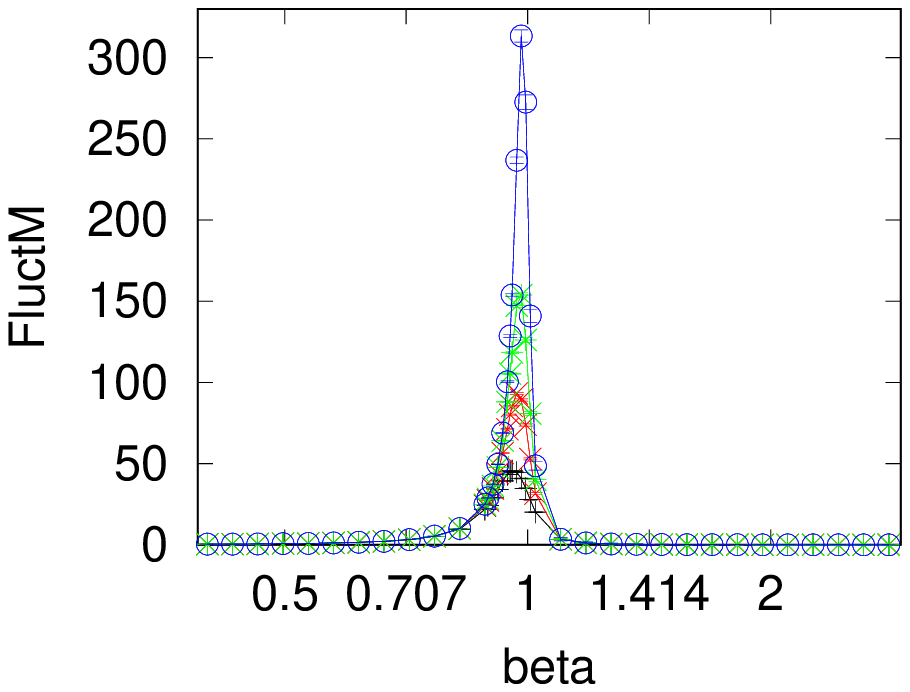}}}}}
\else
\includegraphics[width=8.9cm]{Fig14.ps}
\fi
\caption{(Color online) Disorder fluctuations of magnetization $\chi_1$ of the Ising model ($q=2$),
with a disorder strength $r=3$ and a correlation exponent $a=0.4$ ($y=0.75$),
versus the inverse temperature $\beta=1/k_BT$. In the inset, disorder
fluctuations in the case of uncorrelated disorder.
}\label{FluctM-q_2}
\end{figure}

\begin{figure}
\centering
\ifpsfrag
\psfrag{FluctM}[Bc][Bc][1][1]{$\chi_1$}
\psfrag{beta}[tc][tc][1][0]{$\beta$}
\psfrag{L=32}[Bc][Bc][1][0]{\tiny $L=32$}
\psfrag{L=64}[Bc][Bc][1][0]{\tiny $L=64$}
\psfrag{L=48}[Bc][Bc][1][0]{\tiny $L=48$}
\psfrag{L=96}[Bc][Bc][1][0]{\tiny $L=96$}
\makebox[8.9cm][r]{
\setbox1=\hbox{\includegraphics[height=8.5cm]{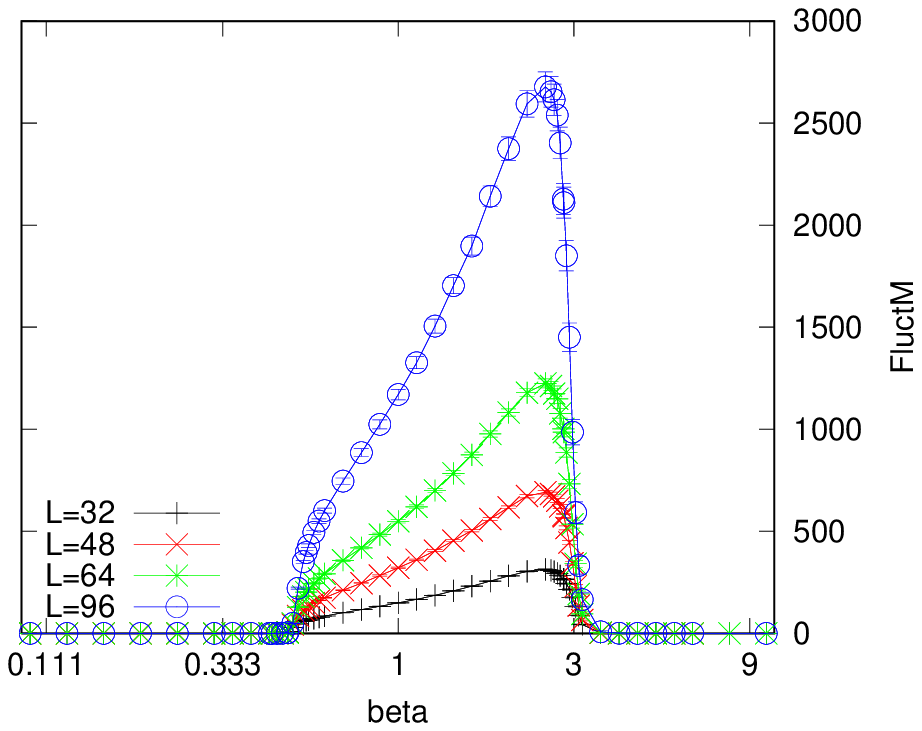}}
\includegraphics[width=8.5cm]{FluctM-q_8-r_7.5-y_0.75.eps}
\llap{\makebox[\wd1][l]{\raisebox{3cm}{\hskip 2.5cm
\includegraphics[height=2.75cm]{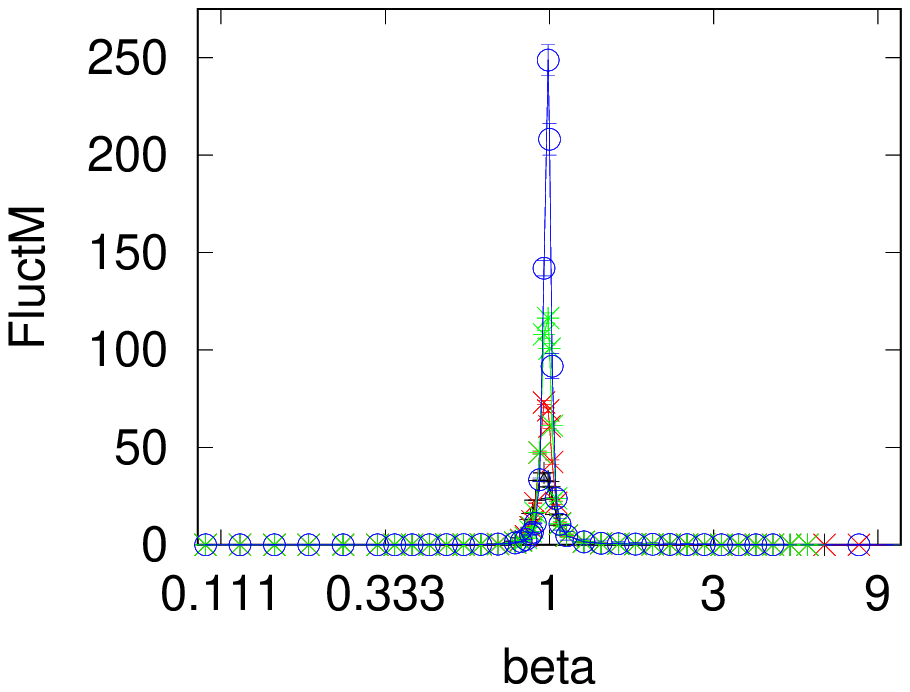}}}}}
\else
\includegraphics[width=8.9cm]{Fig15.ps}
\fi
\caption{(Color online) Disorder fluctuations of magnetization $\chi_1$ of the 8-state Potts model,
with a disorder strength $r=7.5$ and a correlation exponent $a=0.4$ ($y=0.75$),
versus the inverse temperature $\beta=1/k_BT$. In the inset, disorder
fluctuations in the case of uncorrelated disorder.
}\label{FluctM-q_8}
\end{figure}

As pointed out in the previous section, the average susceptibility diverges
with an exponent $\gamma/\nu$ which is close, but not perfectly compatible,
with the hyperscaling relation and the magnetization exponent $\beta/\nu$.
As pointed out in Ref~\cite{EPL}, this violation of hyperscaling is the
result of large disorder fluctuations, as in the 3D random-field Ising
model~\cite{SchwartzSoffel} (RFIM).
We briefly describe in the following the arguments of~\cite{EPL}.
The average magnetic susceptibility, as given by the derivative of the
average magnetization with respect to an external magnetic field, can
be decomposed as
   \ba\overline\chi&=&\beta L^d\big[\overline{\langle m^2\rangle}
     -\overline{\langle m^2\rangle}\big]\nonumber\\
   &=&\underbrace{\beta L^d\big[\overline{\langle m^2\rangle}
     -\overline{\langle m\rangle}^2\big]}_{=\chi_1}-\underbrace{\beta L^d\big[
     \overline{\langle m\rangle^2}-\overline{\langle m\rangle}^2\big]}_{=\chi_2}
   \label{Susc}\ea
The two terms, denoted $\chi_1$ and $\chi_2$, scale differently from the average
susceptibility, i.e. their difference. The second term $\chi_2$ is the numerator
of the ratio $R_m$, defined by Eq. \ref{Rm}. Because the latter tends
to a finite constant in the Griffiths phase (see Fig.~\ref{Rm}),
$\overline{\langle m\rangle^2}-\overline{\langle m\rangle}^2$ scales
as $\overline{\langle m\rangle}^2$, i.e. as $L^{-2\beta/\nu}$.
Therefore, including the $L^d$ prefactor, the susceptibility $\chi_2$
scales as $L^{d-2\beta/\nu}$, i.e. precisely as predicted by the hyperscaling
relation Eq. \ref{Hyperscaling}. The numerical study of the Finite-Size
Scaling of $\chi_1$ reveals that it displays the same scaling behavior.
As can be seen on Fig.~\ref{FluctM-q_2} in the case of the Ising model
and Fig.~\ref{FluctM-q_8}
for the 8-state Potts model, $\chi_1$ is very different from the average
susceptibility. Instead of two peaks separating a region of divergent
susceptibility, a single broader peak is observed. Uncorrelated disorder
leads to a thiner and thiner peak, like the average susceptibility. 
For all temperatures in the Griffiths phase, $\chi_1$ diverges
algebraically with an exponent that will be denoted $(\gamma/\nu)^*$
in the following~\cite{RemarkDroplet}. The exponent $(\gamma/\nu)^*$ is
estimated by following the same procedure as in the previous section.
Effective exponents are first extracted by varying the fitting window.
In contrast to the average magnetic susceptibility, $\chi_1$ displays
relatively weak scaling corrections so the effective exponents can
safely be extrapolated with a linear fit. The extrapolated exponents
$(\gamma/\nu)^*$ are presented in Tables~\ref{Tab4} and \ref{Tab10}.
$(\gamma/\nu)^*$ depends on the disorder correlation exponent $a$
but not on the number of states $q$, nor the strength of disorder $r$.
The temperature dependence in the Griffiths phase is clearly seen
for the largest values of $y$. As claimed above, the exponents are
compatible with the hyperscaling relation and the estimates of
$\beta/\nu$ for any Potts model and at any temperature in the
Griffiths phase.

\begin{table*}
\caption{\label{Tab4}Critical exponent $(\gamma/\nu)^*$ extrapolated from the
Finite-Size Scaling of the quantity $\chi_1$ at the self-dual point $\beta_c=1$.}
\begin{ruledtabular}
\begin{tabular}{l|llllll|l}
$y$ & $0$ & $0.25$ & $0.5$ & $0.75$ & $1$ & $1.25$ & Uncorr. Dis. \\
\hline
$q=2$, $r=2$ & $1.92(4)$ & $1.91(4)$ & $1.90(4)$ & $1.88(4)$ & $1.86(4)$ & $1.83(4)$ & $1.751(13)$\\
$q=2$, $r=3$ & $1.91(3)$ & $1.91(3)$ & $1.89(3)$ & $1.87(3)$ & $1.85(3)$ & $1.81(4)$ & $1.75(2)$\\
\hline
$q=4$, $r=4$ & $1.90(3)$ & $1.89(3)$ & $1.88(3)$ & $1.86(3)$ & $1.83(3)$ & $1.79(3)$ & $1.73(2)$\\
\hline
$q=8$, $r=6$ & $1.90(2)$ & $1.89(2)$ & $1.87(2)$ & $1.86(2)$ & $1.82(3)$ & $1.78(2)$ & $1.72(4)$\\
$q=8$, $r=7.5$ & $1.90(2)$ & $1.90(2)$ & $1.88(2)$ & $1.85(2)$ & $1.82(3)$ & $1.79(3)$ & $1.73(4)$\\
$q=8$, $r=9$ & $1.91(2)$ & $1.89(2)$ & $1.88(2)$ & $1.85(2)$ & $1.83(3)$ & $1.79(3)$ & $1.75(4)$\\
\hline
$q=16$, $r=10$ & $1.90(2)$ & $1.89(2)$ & $1.88(2)$ & $1.86(2)$ & $1.83(2)$ & $1.78(3)$ & $1.74(4)$\\
\end{tabular}
\end{ruledtabular}
\end{table*}

\begin{table*}
\caption{\label{Tab10}Critical exponent $(\gamma/\nu)^*$ extrapolated from the
Finite-Size Scaling of the quantity $\chi_1$ at different temperatures in
the Griffiths phase.}
\begin{ruledtabular}
\begin{tabular}{l|llllll}
$y$ & $0$ & $0.25$ & $0.5$ & $0.75$ & $1$ & $1.25$ \\
\hline
$q=2$, $\beta=1$ & $1.91(3)$ & $1.91(3)$ & $1.89(3)$ & $1.87(3)$ & $1.85(3)$ & $1.81(4)$ \\
$q=2$, $\beta=1.2$ & $1.96(3)$ & $1.95(4)$ & $1.93(4)$ & $1.92(4)$ & $1.92(4)$ & $1.79(6)$\\
\hline
$q=8$, $\beta=0.75$ & $1.88(2)$ & $1.87(2)$ & $1.85(2)$ & $1.82(2)$ & $1.77(2)$ & $1.72(2)$\\
$q=8$, $\beta=1$ & $1.90(2)$ & $1.90(2)$ & $1.88(2)$ & $1.85(2)$ & $1.82(3)$ & $1.79(3)$  \\
$q=8$, $\beta=1.5$ & $1.92(3)$ & $1.92(3)$ & $1.90(2)$ & $1.89(2)$ & $1.87(3)$ & $1.84(3)$ \\
$q=8$, $\beta=2$ & $1.94(3)$ & $1.94(3)$ & $1.92(3)$ & $1.91(3)$ & $1.90(3)$ & $1.78(5)$ \\
\end{tabular}
\end{ruledtabular}
\end{table*}

\begin{figure}
\centering
\ifpsfrag
\psfrag{1/L}[Bc][Bc][1][1]{$1/L$}
\psfrag{Ratio}[Bc][Bc][1][0]{$\chi_1/\chi_2$}
\psfrag{y=0.00}[Bc][Bc][1][0]{\tiny $y=0.00$}
\psfrag{y=0.25}[Bc][Bc][1][0]{\tiny $y=0.25$}
\psfrag{y=0.50}[Bc][Bc][1][0]{\tiny $y=0.50$}
\psfrag{y=0.75}[Bc][Bc][1][0]{\tiny $y=0.75$}
\psfrag{y=1.00}[Bc][Bc][1][0]{\tiny $y=1.00$}
\psfrag{y=1.25}[Bc][Bc][1][0]{\tiny $y=1.25$}
\psfrag{Uncorr.}[Bc][Bc][1][0]{\tiny Uncorr}
\psfrag{Beta=0.75}[Bc][Bc][1][0]{\tiny $\beta=0.75$}
\psfrag{Betac}[Bc][Bl][1][0]{\tiny $\beta=1$}
\psfrag{Beta=1.5}[Bc][Bc][1][0]{\tiny $\beta=1.5$}
\psfrag{Beta=2}[Bc][Bc][1][0]{\tiny $\beta=2$}
\includegraphics[width=4.25cm]{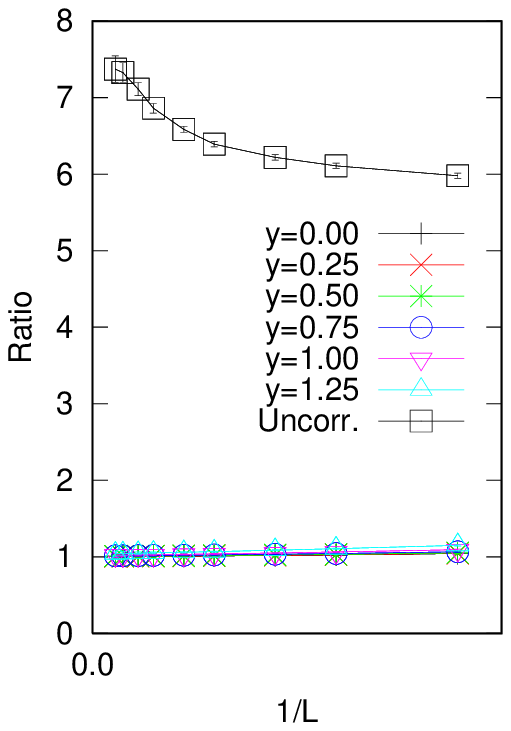}
\includegraphics[width=4.25cm]{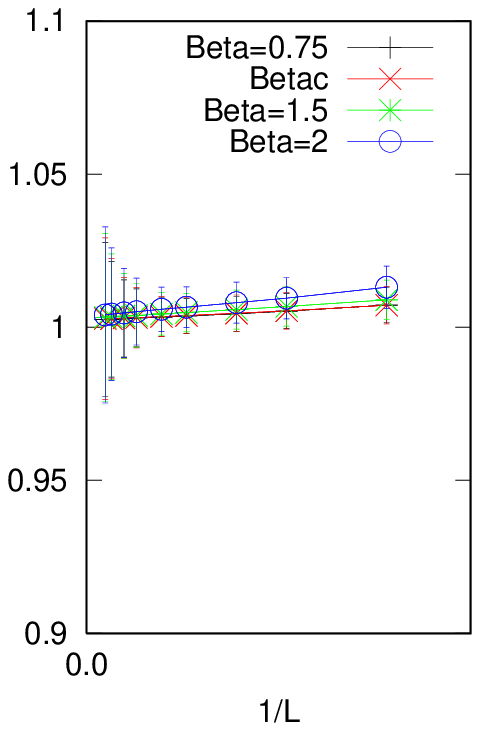}
\else
\includegraphics[width=8.9cm]{Fig16.ps}
\fi
\caption{(Color online) Ratio $\chi_1/\chi_2$ versus the inverse $1/L$ of the lattice size
for various parameters $y$ in the case of the Ising model ($r=3$) on the left
and for several temperatures in the Griffiths phase of the 8-state Potts
($y=0.75$ and $r=7.5$) on the right.}
\label{Ratio}
\end{figure}

Not only $\chi_1$ and $\chi_2$ have the same scaling behavior but their
dominant scaling terms also have the same amplitude $A$, i.e.
  \be \chi_i=AL^{d-2\beta/\nu}\big(1+B_iL^{-\omega_i}+\ldots\big),\quad (i=1,2)
  \label{EqualAmpl}.\ee
As a consequence, their difference behaves at large lattice sizes as
   \be\overline\chi=\chi_1-\chi_2\sim AB_1L^{d-2\beta/\nu-\omega_1}
   -AB_2L^{d-2\beta/\nu-\omega_2}\ee
i.e. with a scaling dimension $\gamma/\nu=d-2\beta/\nu-\min(\omega_1,\omega_2)$.
Our estimates of $\gamma/\nu$, as given in Tables~\ref{Tab3b} and \ref{Tab9b},
and the analysis of the ratio $\overline\chi/L^d\overline{\langle m\rangle}^2$
(see for example Fig.~\ref{Violation-vs-L}) indicate that the
deviation from the hyperscaling relation, i.e. $\min(\omega_1,\omega_2)$, is
small (of order $0.1-0.3$). The only possibility for a restoration of
hyperscaling at large lattice sizes is that the largest correction
of either $\chi_1$ or $\chi_2$ diverge logarithmically, i.e. $\omega_1=0$
or $\omega_2=0$. In any other cases, one should expect a violation of
hyperscaling. Note that the average susceptibility $\overline\chi$
is two to three orders of magnitude smaller than both $\chi_1$ and $\chi_2$,
which means that $B_i\simeq 10^{-2}$. This is in agreement with the observation
that $\chi_1$ and $\chi_2$ only display weak scaling corrections. To check that
both $\chi_1$ and $\chi_2$ have the same dominant amplitude $A$, their ratio
was analyzed. Two particular cases are presented on Fig.~\ref{Ratio}. On
the left, the ratio $\chi_1/\chi_2$ for the Ising model goes to the expected
limit $1$ for all parameters $y$ considered while for uncorrelated disorder,
a very different limit is observed. On the right, this ratio is plotted
in the case of the 8-state Potts at several temperatures in the Griffiths
phase. Again, the data goes to the limit $1$. The same analysis, reproduced
for all numbers of states $q$, strength of disorder $r$, or temperatures
in the Griffiths phase, leads to the same conclusion.
In all cases, the leading amplitudes of $\chi_1$ and
$\chi_2$ are shown to be identical. Consequently, the dominant contributions
of $\chi_1$ and $\chi_2$ cancel out and the hyperscaling relation is broken in
the entire Griffiths phase. In the case of uncorrelated disorder, the ratio
goes to a value very different from 1 (see left of Fig.~\ref{Ratio}).
Therefore, the dominant contributions of $\chi_1$ and $\chi_2$ do not
cancel out in this case and the hyperscaling relation is not broken.

The same analysis can be performed in the energy sector. However, as already
mentioned, it is not possible to decide whether the hyperscaling relation
${\alpha\over\nu}={2\over\nu}-d$ is broken or not because the specific heat does
not diverge and therefore its singular part cannot be separated from the
regular background. Nevertheless, one can check that the same mechanism is
present, which implies that hyperscaling is broken unless the first correction
is only logarithmic. The average specific heat is decomposed as
   \ba\overline C&=&\beta^2L^d\big[\overline{\langle e^2\rangle}
     -\overline{\langle e\rangle^2}\big]\nonumber\\
   &=&\underbrace{\beta^2L^d\big[\overline{\langle e^2\rangle}
     -\overline{\langle e\rangle}^2\big]}_{=C_1}-\underbrace{\beta^2L^d\big[
     \overline{\langle e\rangle^2}-\overline{\langle e\rangle}^2\big]}_{=C_2}.
   \label{SpecHeat}\ea
The scaling behavior of $C_2$ cannot be deduced from the ratio $R_e$ defined
by Eq.~\ref{Re}. In contrast to the magnetic sector, $R_e$ vanishes in the
thermodynamic limit, meaning that the energy density becomes a self-averaging
quantity~\cite{RemarkSelfAvgE}.
Both $C_1$ and $C_2$ display a behavior with temperature very
different from the average specific heat. In contrast to $\overline C$, they
diverge in the Griffiths phase and, because of the $\beta^2$ prefactor, they
keep on growing as the temperature is decreased. We have tested the Finite-Size
Scaling of $C_1$ at several temperatures in the Griffiths phase. While
the specific heat is almost independent of the lattice size, $C_1$ diverges
algebraically with a large exponent $(\alpha/\nu)^*$. As shown in
Tables~\ref{Tab5} and \ref{Tab11}, $(\alpha/\nu)^*$ depends only on the
disorder correlation exponent $a$, and therefore on $y$, but not on
the number of states $q$, the strength of disorder $r$, nor the temperature.
The estimates are in good agreement with $d-a$ which means that the energy
fluctuations are controlled by the disorder fluctuations. This result
implies that a strong coupling $J_2$ essentially freezes the relative state
of the two spins, while a weak one $J_1$ leads to an irrelevant constraint
between them. Amazingly, the ratio $C_1/C_2$ goes to a constant in excellent
agreement with 1 for all temperatures in the Griffiths phase. As a consequence,
the dominant contributions of $C_1$ and $C_2$ cancel out exactly, leading to
an algebraic behavior of the specific heat with a much smaller exponent
$\alpha/\nu \ll (\alpha/\nu)^*$ than both $C_1$ and $C_2$. Since $\alpha/\nu
\le 0$, the largest scaling correction of $C_1$ and $C_2$ decays faster
than $L^{-(\alpha/\nu)^*}$. In the case of uncorrelated disorder,
the ratio $C_1/C_2$ goes to a value different from $1$, even though error
bars increase rapidly at large lattice sizes and finally include 1. The
cancellation does not take place and the equality $\alpha/\nu=(\alpha/\nu)^*$
is expected.
\\

In the magnetic sector, the exponent $(\gamma/\nu)^*$ was shown to satisfy
the hyperscaling relation $(\gamma/\nu)^*=d-2\beta/\nu=d-2x_\sigma$
with the exponent $\beta/\nu$. Assuming that this is also the case in the
energy sector, $(\alpha/\nu)^*$ is conjectured to satisfy the hyperscaling
relation $(\alpha/\nu)^*=d-2x_\varepsilon$ where $x_\varepsilon$ is the
energy scaling dimension. As discussed above, $(\alpha/\nu)^*$ is compatible
with $d-a$ which implies $x_\varepsilon=a/2$, i.e. the energy-energy correlation
functions, decaying as $r^{-2x_\varepsilon}$, are determined by the coupling
correlations in the Griffiths phase.

\begin{table*}
\caption{\label{Tab5}Critical exponent $(\alpha/\nu)^*$ estimated from the
Finite-Size Scaling of the quantity $C_1$ at the self-dual point $\beta_c=1$.
No extrapolation is needed in this case.}
\begin{ruledtabular}
\begin{tabular}{l|llllll|l}
$y$ & $0$ & $0.25$ & $0.5$ & $0.75$ & $1$ & $1.25$ & Uncorr. Dis. \\
\hline
$q=2$, $r=2$ & $1.75(2)$ & $1.73(2)$ & $1.68(2)$ & $1.61(2)$ & $1.51(2)$ & $1.35(3)$ & $0.23(6)$\\
$q=2$, $r=3$ & $1.748(11)$ & $1.730(12)$ & $1.681(13)$ & $1.608(14)$ & $1.506(15)$ & $1.35(2)$ & $0.18(6)$\\
\hline
$q=4$, $r=4$ & $1.749(9)$ & $1.731(10)$ & $1.682(10)$ & $1.610(11)$ & $1.507(12)$ & $1.34(2)$ & $0.28(5)$\\
\hline
$q=8$, $r=6$ & $1.749(8)$ & $1.729(9)$ & $1.681(9)$ & $1.609(9)$ & $1.506(12)$ & $1.343(14)$ & $0.26(5)$\\
$q=8$, $r=7.5$ & $1.748(8)$ & $1.730(8)$ & $1.681(9)$ & $1.608(10)$ & $1.505(11)$ & $1.344(14)$ & $0.21(5)$\\
$q=8$, $r=9$ & $1.747(8)$ & $1.729(8)$ & $1.680(9)$ & $1.607(10)$ & $1.503(11)$ & $1.340(13)$ & $0.18(5)$\\
\hline
$q=16$, $r=10$ & $1.747(8)$ & $1.731(9)$ & $1.681(9)$ & $1.606(9)$ & $1.502(10)$ & $1.341(13)$ & $0.20(5)$\\
\hline
$d-a$ & $1.75$ & $1.714$ & $1.667$ & $1.600$ & $1.500$ & $1.333$  & 0 \\
\end{tabular}
\end{ruledtabular}
\end{table*}

\begin{table*}
\caption{\label{Tab11}Critical exponent $(\alpha/\nu)^*$ estimated from the
Finite-Size Scaling of the quantity $C_1$ at different temperatures in
the Griffiths phase. No extrapolation is needed in this case.
}
\begin{ruledtabular}
\begin{tabular}{l|llllll}
$y$ & $0$ & $0.25$ & $0.5$ & $0.75$ & $1$ & $1.25$ \\
\hline
$q=2$, $\beta=1$ & $1.748(11)$ & $1.730(12)$ & $1.681(13)$ & $1.608(14)$ & $1.506(15)$ & $1.35(2)$ \\
$q=2$, $\beta=1.2$ & $1.748(12)$ & $1.730(12)$ & $1.682(13)$ & $1.606(14)$ & $1.51(2)$ & $1.34(2)$\\
\hline
$q=8$, $\beta=0.75$ & $1.749(8)$ & $1.731(8)$ & $1.684(9)$ & $1.609(10)$ & $1.506(10)$ & $1.342(13)$\\
$q=8$, $\beta=1$ & $1.748(8)$ & $1.730(8)$ & $1.681(9)$ & $1.608(10)$ & $1.505(11)$ & $1.344(14)$  \\
$q=8$, $\beta=1.5$ & $1.747(8)$ & $1.729(9)$ & $1.680(9)$ & $1.605(10)$ & $1.503(11)$ & $1.338(14)$ \\
$q=8$, $\beta=2$ & $1.747(8)$ & $1.730(9)$ & $1.680(10)$ & $1.605(10)$ & $1.503(12)$ & $1.340(14)$ \\
\end{tabular}
\end{ruledtabular}
\end{table*}

The same procedure is applied to the quantity
     \ba -{d\over d\beta}\ln\overline{\langle m\rangle}
     &=&L^d{\overline{\langle me\rangle-\langle m\rangle\langle e\rangle}
       \over\overline{\langle m\rangle}}\\
     =&&\underbrace{L^d{\overline{\langle me\rangle}-\overline{\langle m\rangle}
       \ \overline{\langle e\rangle}\over\overline{\langle m\rangle}}}_{=X_1}
     -\underbrace{L^d{\overline{\langle m\rangle\langle e\rangle}
         -\overline{\langle m\rangle}\ \overline{\langle e\rangle}
         \over\overline{\langle m\rangle}}}_{=X_2}
     \nonumber\ea
Both terms in the second line diverge algebraically in the Griffiths phase.
Like the specific heat, the exponent $1/\nu^*$ extracted from this divergence
does not depend on the number of states $q$, the strength of disorder $r$,
nor the temperature but only on disorder correlations, i.e. on $y$ or
equivalently on $a$. As can be observed in Tab.~\ref{Tab7} and \ref{Tab13},
the numerical estimates are compatible with $1/\nu^*=d-a/2$. Interestingly,
the exponents $(\alpha/\nu)^*$ and $1/\nu^*$ are compatible with the
hyperscaling relation
    \be\left({\alpha\over\nu}\right)^*={2\over\nu^*}-d.\ee
This relation is of course exactly satisfied by the two conjectures
$(\alpha/\nu)^*=d-a$ et $1/\nu^*=d-a/2$. Assuming $1/\nu^*=d-x_\varepsilon$,
the same value of the energy scaling dimension, i.e. $x_\varepsilon=a/2$,
is obtained. The ratio $X_1/X_2$ is compatible with the value
$1$ at all temperatures in the Griffiths phase. Therefore, the dominant
contributions of $X_1$ and $X_2$ cancel out, which explains the very different
exponent $1/\nu\ll 1/\nu^*$ with which the difference $-{d\over d\beta}
\ln\overline{\langle m\rangle}=X_1-X_2$ diverge. We note that the correction
exponent $\omega=1/\nu^*-1/\nu$ that is responsible for such a large
difference is not an integer, i.e. the largest correction is not
analytic, and depends on the disorder correlation exponent $a$.
It is however smaller than for the specific heat.

\begin{table*}
\caption{\label{Tab7}Critical exponent $1/\nu^*$ extrapolated from the
Finite-Size Scaling of the quantity $X_1$ at the self-dual point $\beta_c=1$.}
\begin{ruledtabular}
\begin{tabular}{l|llllll|l}
$y$ & $0$ & $0.25$ & $0.5$ & $0.75$ & $1$ & $1.25$ & Uncorr. Dis. \\
\hline
$q=2$, $r=2$ & $1.88(5)$ & $1.87(6)$ & $1.85(6)$ & $1.82(5)$ & $1.77(6)$ & $1.70(7)$ & $1.00(7)$\\
$q=2$, $r=3$ & $1.88(4)$ & $1.87(4)$ & $1.84(4)$ & $1.82(4)$ & $1.76(5)$ & $1.68(6)$ & $0.97(10)$\\
\hline
$q=4$, $r=4$ & $1.88(3)$ & $1.86(3)$ & $1.84(3)$ & $1.80(3)$ & $1.75(3)$ & $1.66(4)$ & $1.08(10)$\\
\hline
$q=8$, $r=6$ & $1.88(3)$ & $1.86(3)$ & $1.84(3)$ & $1.81(3)$ & $1.75(3)$ & $1.66(4)$ & $1.10(10)$\\
$q=8$, $r=7.5$ & $1.88(2)$ & $1.87(2)$ & $1.84(2)$ & $1.80(3)$ & $1.75(3)$ & $1.66(4)$ & $1.09(10)$\\
$q=8$, $r=9$ & $1.88(2)$ & $1.87(3)$ & $1.84(3)$ & $1.80(3)$ & $1.75(3)$ & $1.67(4)$ & $1.08(10)$\\
\hline
$q=16$, $r=10$ & $1.87(3)$ & $1.86(2)$ & $1.84(2)$ & $1.80(3)$ & $1.75(3)$ & $1.66(3)$ & $1.09(10)$\\
\hline
$d-a/2$ & $1.875$ & $1.857$ & $1.835$ & $1.8$ & $1.75$ & $1.667$ & $1$ \\
\end{tabular}
\end{ruledtabular}
\end{table*}

\begin{table*}
\caption{\label{Tab13}Critical exponent $1/\nu^*$ extrapolated from the
Finite-Size Scaling of the quantity $X_1$ at different temperatures in the Griffiths phase.}
\begin{ruledtabular}
\begin{tabular}{l|llllll}
$y$ & $0$ & $0.25$ & $0.5$ & $0.75$ & $1$ & $1.25$ \\
\hline
$q=2$, $\beta=1$ & $1.88(4)$ & $1.87(4)$ & $1.84(4)$ & $1.82(4)$ & $1.76(5)$ & $1.68(6)$ \\
$q=2$, $\beta=1.2$ & $1.91(4)$ & $1.90(4)$ & $1.86(5)$ & $1.82(5)$ & $1.79(5)$ & $1.61(7)$\\
\hline
$q=8$, $\beta=0.75$ & $1.86(3)$ & $1.86(2)$ & $1.83(3)$ & $1.80(2)$ & $1.74(3)$ & $1.66(3)$\\
$q=8$, $\beta=1$ & $1.88(2)$ & $1.87(2)$ & $1.84(2)$ & $1.80(3)$ & $1.75(3)$ & $1.66(4)$  \\
$q=8$, $\beta=1.5$ & $1.89(3)$ & $1.88(3)$ & $1.85(3)$ & $1.82(3)$ & $1.76(3)$ & $1.67(4)$ \\
$q=8$, $\beta=2$ & $1.89(3)$ & $1.88(3)$ & $1.86(3)$ & $1.82(3)$ & $1.77(4)$ & $1.60(5)$ \\
\end{tabular}
\end{ruledtabular}
\end{table*}

\section{\label{Sec6}Conclusions}
A Potts model with algebraically-decaying coupling correlations is studied by
large-scale Monte Carlo simulations. Such a disorder is obtained by coupling
the polarization density of a quenched self-dual Ashkin-Teller model to the
energy density of the Potts model. By construction, the disorder is not
generated by a Gaussian action and therefore, multiple-point correlation
functions are not trivially given by the Wick theorem. As a consequence,
the model is outside of the scope considered by Weinrib and Halperin in the
case of the $O(n)$-model. Our model shares two important similarities with
the McCoy-Wu model: a Griffiths phase occurs in a finite range of temperatures
around the self-dual point and scaling dimensions are independent of the
number of Potts states $q$. This is probably a general feature of random
systems with sufficiently strong disorder correlations.

In contrast to energy, magnetization is shown
to be non self-averaging in the Griffiths phase. Magnetization and magnetic
susceptibility display algebraic behaviors with the lattice size at all
temperatures in the phase Griffiths phase. The exponent $\beta/\nu$
does not depend on the number of Potts states $q$ nor the strength of disorder
$r$ but varies with the disorder correlation exponent $a$ and the
temperature in the Griffiths phase. Our estimates of $\gamma/\nu$
display a small violation of the hyperscaling relation. This violation is shown
to be caused by the exact cancellation of two terms, $\chi_1$ and $\chi_2$,
whose difference gives the average susceptibility, as in the 3D
Random-Field Ising model. Such a mechanism leads to an hyperscaling violation
unless the largest correction of any of the two terms $\chi_1$ or $\chi_2$
diverge only logarithmically. From the scaling of $\chi_1$ and $\chi_2$,
an exponent $(\gamma/\nu)^*$ satisfying the hyperscaling relation can be
extracted. Because the specific heat does not diverge, the exponent
$\alpha/\nu$ is negative or zero. However, it can also be written as the
difference of two diverging terms. From them, an exponent $(\alpha/\nu)^*$
is defined and shown to be compatible with $d-a$ for any number of Potts
state and any temperature in the Griffiths phase. Because the same mechanism
than in the magnetic sector takes place, the energy scaling dimension is
conjectured to be given by $(\alpha/\nu)^*=d-2x_\varepsilon$ which implies
$x_\varepsilon=a/2$. The exponent $\nu$ is extracted from the Finite-Size
Scaling of ${d\over d\beta}\overline{\langle m\rangle}$ at different
temperatures of the Griffiths phase. Again, this quantity can be written
as a difference of two terms, diverging with a much larger exponent
$1/\nu^*$ compatible with $d-a/2$, again for any number of Potts states
and any temperature in the Griffiths phase.

Of course, these results have been obtained for finite-size systems
so we cannot exclude completely the possibility that the Griffiths
phase disappears at much larger lattice sizes and that the hyperscaling
relation is restored. Indeed, one could argue that the existence of
this Griffiths phase is related to the large fluctuations of the number
of strong and weak couplings from sample to sample. These fluctuations
are indeed expected to vanish in the thermodynamic limit as $L^{-a/2}$
but are still very large for the lattice sizes that were considered,
much larger than for uncorrelated disorder. If the Griffiths phase
is only due to these fluctuations, its width should also vanish
in the thermodynamic limit as $L^{-a/2}$. The numerical data
seems to indicate that it is not the case: the distance between the two
peaks of the average magnetic susceptibility on figure~\ref{chi-q_8}
is roughly $2.44$ for $L=32$ and $2.34$ for $L=96$. There is therefore
a reduction of the width but much smaller that the expected factor
$(32/96)^{-0.2}\simeq 1.25$ for $y=0.75$ ($a=0.4$). The Griffiths phase
does not seem to be a consequence of only the fluctuations of the number
of strong and weak couplings. Of course, much larger lattice sizes would help
to clarify this point. Another possibility would be to study correlated
disorder with a larger exponent $a$, which is not possible with the
two-dimensional Ashkin-Teller model.

Another interesting point would be to understand precisely why the
critical behavior observed in this model does not fall into the same
universality as the $\phi^4$ model studied by Weinrib and Halperin.
Further Renormalization Group studies could probably clarify the role
of $n$-point disorder correlations functions.

\begin{acknowledgments}
The author is grateful to Sreedhar Dutta for warm discussions
and to the Indian Institute for Science Education and Research
(IISER) of Thiruvananthapuram where part of this work was done.
The author also thanks Francesco Parisen Toldin for having pointed
out the fact that $R_\chi$ decreases with the lattice size $L$ as
$1/\ln L$ in the case of the Ising model with uncorrelated disorder.
\end{acknowledgments}


\begin{thebibliography}{99}
\bibitem{Remark1}
Errors on the derivative ${d\over d\beta}\ln\overline{\langle m\rangle}$
were considerably improved compared to Ref.~\cite{EPL}. As discussed in the
latter reference, this derivative is computed as the difference of two
observables, $\overline{\langle me\rangle}$ and $\overline{\langle m\rangle
\langle e\rangle}$, which are several orders of magnitude larger than their
difference. If these two observables are computed separately, as it was the
case in Ref.~\cite{EPL}, it is natural to estimate the error on their
difference as the sum of their respective errors. However, this does not
reflect the true error on the difference $\bar X=\overline{\langle me\rangle}
-\overline{\langle m \rangle\langle e\rangle}$ because the two are
correlated. The difference was therefore computed during the simulation
and the error, estimated by $\sqrt{(\overline{X^2}-\overline{X}^2)/N}$
where $N$ is the number of disorder configuration.
\bibitem{NuLarge}
The large estimates of the exponent $\nu$ given in the
text contrast with the values of this exponent for the (uncorrelated)
random Potts models, known to be close to $1$. However, large
exponents $\nu$ are frequent. In some cases, the correlation length
exponent even goes to infinity, which means that the correlation length
grows exponentially fast as the critical temperature is approached.
It is the case of the 1D Ising model for which $\xi\sim e^{2/T}$ and
of the anti-ferromagnetic Ising model on a triangular lattice.
In the neighborhood of a Kosterlitz-Thouless phase transition, like
in the 2D XY model for example, the correlation length behaves as
$\xi\sim e^{at^{-\sigma}}$ where $t=T-T_{\rm KT}$. An algebraic growth
with an exponent $\nu\simeq 10-40$ is therefore not a particularly
fast growth.
Moreover, note that the Weinrib-Halperin prediction $\nu=2/a$
gives also large values (between $3$ and $8$) for the disorder
correlations considered. Finally, note that, as commented
at the beginning of Sec.~\ref{Sec4}, the exponent $\nu$ is defined
only through Finite-Size Scaling so a scaling law such as $\xi\sim
t^{-\nu}$ may not be applicable in the Griffiths phase if
the correlation length $\xi$ diverges at any temperature
in this region.
\bibitem{RemarkDroplet}
In the context of the 3D RFIM, the susceptibility $\chi_1$ was also
shown to diverge at the critical point with a different exponent than
the average susceptibility~\cite{BrayMoore}. Different notations were
introduced. The correspondence with our notation is $(\gamma/\nu)^*
=4-\bar\eta$.
\bibitem{RemarkSelfAvgE}
The regular part of the energy density is responsible for the fact that
the latter is a self-averaging quantity. According to the Finite-Size
Scaling behaviors discussed in the text, the ratio $R_e$ decays
algebraically with the lattice size:
   \be R_e={\overline{\langle e\rangle^2}-\overline{\langle e\rangle}^2
    \over\overline{\langle e\rangle}^2}
    =\beta^2{C_2/L^d\over\overline{\langle e\rangle}^2}
    \sim L^{(\alpha/\nu)^*-d}\ee
since $(\alpha/\nu)^*<d$ and $\overline{\langle e\rangle}\sim L^0$.
In the magnetic sector, $R_m={\chi_2/L^d\over
\overline{\langle m\rangle}^2}$. In contrast to $\overline{\langle e\rangle}^2$,
the denominator is singular and scales as $L^{-2\beta/\nu}$.
Because $\chi_2\sim L^{(\gamma/\nu)^*}$ where $(\gamma/\nu)^*$ satisfies the
hyperscaling relation, $R_m$ scales as $L^0$, which allows for a non-vanishing
limit and therefore a possible absence of self-averaging.
\end{thebibliography}
\end{document}